\documentclass[%
reprint,
amsmath,
amssymb,
aip,
jcp,
]{revtex4-2}
\usepackage{graphicx}% Include figure files
\usepackage{bm}% bold math
\usepackage[hidelinks]{hyperref}% add hypertext capabilities 
\usepackage{dcolumn}% Align table columns on decimal point
\usepackage{booktabs}           % makes tables look good 

\usepackage{array}
\newcolumntype{L}[1]{>{\raggedright\arraybackslash}m{#1}}
\newcolumntype{C}[1]{>{\centering\arraybackslash}m{#1}}
\newcolumntype{R}[1]{>{\raggedleft\arraybackslash}m{#1}}

\usepackage[linesnumbered,ruled,commentsnumbered]{algorithm2e}
\usepackage{tikz}
\usetikzlibrary{shapes, arrows.meta, positioning}

\definecolor{Rochester}{RGB}{0,59,113}
\definecolor{dandelion}{RGB}{255,209,0}
\tikzstyle{state}=[circle, draw, fill=Rochester, thick, minimum width=2em]
\tikzstyle{basis}=[circle, draw, fill=dandelion, thick, minimum width=2em]
\tikzstyle{operator}=[rectangle, draw, fill=yellow!50, thick, minimum width=2em, minimum height=2em]
\tikzstyle{other}=[rectangle, draw, fill=green!50, thick, minimum width=2em, minimum height=2em]

\newcommand{\kB}{\ensuremath{k_\mathrm{B}}}
\newcommand{\iu}{\ensuremath{\mathrm{i}}}
\newcommand{\invcm}{\ensuremath{\mathrm{cm}^{-1}}}

\newcommand{\set}[1]{\ensuremath{\left\{{#1}\right\}}}

\usepackage{mathtools}
\DeclarePairedDelimiter\bra{\langle}{\rvert}
\DeclarePairedDelimiter\ket{\lvert}{\rangle}
\DeclarePairedDelimiterX\braket[2]{\langle}{\rangle}{#1\,\delimsize\vert\,\mathopen{}#2}

\newcommand{\ev}[1]{\left\langle {#1} \right\rangle} 
\newcommand{\ip}[2]{\left\langle {#1} \middle| {#2} \right\rangle}

\newcommand{\Tr}{\mathrm{Tr}}
\newcommand{\dv}[1]{\frac{\mathrm{d}}{\mathrm{d}{#1}}}

\newcommand{\dd}[1]{\mathrm{d}{#1}}
\newcommand{\abs}[1]{\left\lvert {#1} \right\rvert}
\newcommand{\qand}{\quad\mathrm{and}\quad}

\newcommand{\comm}[2]{\left[{#1}, {#2}\right]}

\newcommand{\qtya}[1]{{\left( {#1} \right)}}

\renewcommand{\Im}{\operatorname{Im}}

\begin{document}

\title{Comparison Between Explicit and Implicit Discretization Strategies for a Dissipative Thermal Environment}

\author{Xinxian Chen}
 \email{xchen106@ur.rochester.edu}
\affiliation{Department of Chemistry, University of Rochester, Rochester, New York 14627, United States}

\author{Ignacio Franco}
\email{ignacio.franco@rochester.edu (Corresponding author)}
\affiliation{Department of Chemistry, University of Rochester, Rochester, New York 14627, United States}
\affiliation{Department of Physics and Astronomy, University of Rochester, Rochester, New York 14627, United States}
\affiliation{The Institute of Optics, University of Rochester, Rochester, New York 14627, United States}

\date{\today}

\begin{abstract}
  We investigate strategies for simulating open quantum systems coupled to dissipative baths by comparing explicit wavefunction-based discretization (via multi-layer multi-configuration time-dependent Hartree, ML-MCTDH) and implicit density matrix-based master equation method (via tree tensor network hierarchical equations of motion, TTN-HEOM). For dissipative baths characterized by exponentially decaying bath correlation functions, the implicit discretization approach of HEOM---rooted in bath correlation function decompositions---proves significantly more efficient than explicit discretization of the bath into discrete harmonic modes.  Explicit methods like ML-MCTDH, require extensive mode discretization to approximate continuum baths, leading to computational bottlenecks. 
  Case studies for two-level systems and a Fenna--Matthews--Olson complex model highlight TTN-HEOM's superiority in capturing dissipative dynamics with relaxations with a minimal number of auxiliary modes, while  {the} explicit methods are as exact as the HEOM in pure dephasing regimes.  {This comparison is enabled by the \texttt{TENSO}  package  which has both ML-MCTDH and TTN-HEOM implemented using the same  computational structure and propagation strategy.}
  \end{abstract}

\maketitle

\section{Introduction} 

The time-dependent Schr\"odinger equation (TDSE) is one of the most fundamental equations for quantum dynamics.\cite{Tannor2007}
However, its direct application to computationally simulate the quantum dynamics of a system coupled to a macroscopic environment (which is a common physical situation in molecular, material, and quantum information science) is challenging,
because of the large number of degrees of freedom (DoFs) involved.

In the context of open quantum systems, 
it is well-known that any quantum bath can be mapped to a collection of bosons provided the system-bath interaction can be captured to second order in perturbation theory.\cite{Feynman1963, Caldeira1983, Caldeira1993} 
This situation is common for a system in the condensed phase where system-bath interactions are diluted over a macroscopic number of DoFs.\cite{Suarez1991, Makri1995} 
It also includes the case of a bath of independent spins, where a macroscopic spin bath can also be mapped to a macroscopic bosonic bath in the linear response limit.\cite{Hu2010,Kim2024a,Ying2024} 
This feature makes simulations in the presence of a bosonic bath of particular interest.

Computational simulations of such a situation using the TDSE invariably require a discretization of such a bosonic bath using a finite number of modes.
This is needed to explicitly represent the quantum state of the system plus bath as a vector in a finite-dimensional Hilbert space,\cite{Tannor2007}  with each discretized mode representing one DoF of the bath.
The challenge in this strategy is the curse of dimensionality, where the size of the Hilbert space grows exponentially with the number of discretized modes in the bath.

One way to mitigate this challenge is to use a tensor network decomposition of the quantum state in the TDSE, which is a state-of-the-art strategy in many-body physics and chemistry that can be used in both stationary \cite{Schollwoeck2011, White1992, White1993} and time-dependent problems.\cite{Cazalilla2002, Luo2003, Feiguin2005a, Keller2015, Kloss2017, Ren2018} 
In those methods, the many-body wavefunction is decomposed into a tensor network based on sequential singular value decompositions of the quantum mechanical state to reduce the dimensionality of the quantum simulation while still capturing most of essential principal components.
In particular, the multi-configuration time-dependent Hartree (MCTDH) method \cite{Meyer1990} and its multi-layer (ML) extension \cite{Wang2003,Wang2021} are based on a hierarchical Tucker decomposition \cite{Grasedyck2010} of the wavefunction, which has a tensor network with a tree-like structure. 
This wavefunction-based method was originally developed for simulating quantum dynamics at zero temperature.
However, they have also been extended to finite temperature scenarios,\cite{Wang2006, Schollwoeck2011, Ren2018} which requires a discretization strategy to simulate a dissipative and continuous bath \cite{Vega2015,Walters2017}  {that leads to thermalization of the system}.
Since the overall dynamics remains unitary, this strategy can struggle to fully capture the irreversible nature of a truly dissipative bath with infinite DoFs.\cite{Tanimura2020} 

Another strategy to simulate quantum dynamics in a thermal bath are those produced by quantum master equations,\cite{Lidar2020} 
where only the dynamics of the reduced density matrix of the system is propagated and the influence of the bath is captured implicitly.
This avoids the insurmountable computational cost of explicitly propagating the quantum dynamics of a macroscopic bath.
The hierarchical equation of motion (HEOM) theory is one of the most advanced and successful quantum master equations for simulating the open quantum system coupled to a thermal dissipative bath in a numerically exact manner.\cite{Tanimura1990,Tanimura2020} 
However, the standard HEOM theory has been mostly limited to simple bath models because the computational cost for the dynamics grows exponentially with the complexity of the bath and the number of levels in the system.\cite{Shi2009,Shi2018,Yan2021,Borrelli2021} 
Specifically, the interaction between the system and the bath can be characterized by a spectral density $J(\omega)$ which quantifies the frequencies $\omega$ of the bath and its interaction strength with the system.
While spectral densities of relevance in chemistry are highly structured, currently the HEOM calculations are typically limited to simple spectral density models,\cite{Tanimura2020,Ikeda2020,Lindoy2023} in which the spectral density leads to a bath correlation function (BCF) that can be decomposed into a finite number of exponential components.
This has prevented the use of HEOM to understand excited state molecular dynamics in realistic chemical baths. 
To address this, many techniques have been developed.
For instance, one can apply the filtering algorithm based on the fact that the most of the auxiliary density matrices in the HEOM for capturing the influence of the bath are zero or almost zero matrices,\cite{Shi2009} which makes it possible to do HEOM with tens of exponential components in the BCF.
This technique, however, is sensitive to the time step used for the filtering, and is hard to suppress the number of non-zero auxiliary density matrices in some models.\cite{Dan2023} 

To overcome this problem, in Ref.~\onlinecite{Chen2025}, we developed a tree tensor network (TTN) HEOM method that can be used to simulate the open quantum dynamics of a system coupled to a harmonic bath with arbitrary spectral density discretized by a series with finite many terms, where each term corresponds to one DoF for representing the system-bath interaction.
The TTN-HEOM preserves a master equation formalism that admits general numerical techniques for solving ordinary differential equations, and also allows specific step-wise propagators developed from the tensor network and MCTDH communities.
Our efforts augment and complement other strategies to integrate tensor network techniques into the HEOM.\cite{Shi2018,Borrelli2019,Borrelli2021,Yan2021,Li2022,Ke2023} 

In this work, we revisit and compare the performance of TDSE and the HEOM to describe a system interacting with a bosonic bath.
Specifically,
we compare the performance of the direct discretization strategy in the TDSE for a dissipative bath with the implicit discretization strategy in the HEOM.
For practical simulations, we incorporate the tensor network framework in both TDSE and HEOM, namely ML-MCTDH and TTN-HEOM, to simulate the quantum dynamics of a system coupled to a harmonic bath.
Our results suggest that for dissipative baths characterized by exponentially decaying bath correlation functions (BCFs), the implicit discretization strategy is more efficient than the explicit strategy when tensor network techniques are employed in both approaches.

 {This comparison is enabled by the \texttt{TENSO} package,\cite{Chen2025} which offers the general implementation of TTN decomposition for general ordinary differential equations for high-dimensional tensors  with a generator in a sum-of-product form. Both the ML-MCTDH and TTN-HEOM are implemented in \texttt{TENSO} using the same propagation strategy and computational structure. Thus, any differences between the two arise just because of the computational cost of the method and not because of differences in implementation or computational architecture.  
}

This paper is organized as follows.
In Sec.~\ref{sec:ch7-theory}, we briefly introduce the TDSE with an effective thermal bath used for ML-MCTDH, and the HEOM method that can be used with TTN.
In Sec.~\ref{sec:ch7-strategy}, we introduce strategies for discretization of a dissipative bath for the ML-MCTDH.
The decomposition of the BCF is also discussed as an implicit way to discretize the bath.
Section~\ref{sec:ch7-results} demonstrates examples of these methods for both dephasing and dissipative dynamics, and compares the performance of the ML-MCTDH and TTN-HEOM.
Our conclusions are summarized in Sec.~\ref{sec:ch7-con}.

\section{Quantum master equation for a system in a harmonic bath}\label{sec:ch7-theory}

\subsection{Wavefunction-based methods}\label{sec:ch7-wfn-ZT}
For the sake of completeness, in this section we briefly recall the basic theory of wavefunction-based dynamics for a finite-state quantum system coupled to a harmonic bath.
The system-bath interaction is described by the Hamiltonian
\begin{align}\label{eq:ch7-full-h}
  H &= H_{\text{S}}(t) + H_{\text{SB}} + H_{\text{B}},
\end{align}
where $H_{\text{S}}(t)$ is the Hamiltonian of the system of interest,
$H_{\text{B}}$ is the formal Hamiltonian for the whole bath,
and $H_{\text{SB}}$ is the influence of the bath.
For simplicity in presentation, here we focus on the case when $H_{\text{SB}}$ is one bilinear coupling between the system plus bath
\begin{equation}\label{eq:ch7-sb-h}
  H_{\text{SB}} = Q_{\text{S}} \otimes X_{\text{B}},
\end{equation}
where $Q_{\text{S}}$ is an operator of the system, and $X_{\text{B}} $ of the bath.
We further require the bath to be described by a harmonic model with
\begin{align}\label{eq:ch7-bath-h}
  H_{\text{B}} &= \sum_{j} \omega_j  a_{j}^\dagger a_{j},
    \\
    \label{eq:ch7-bath-x}
  X_{\text{B}} &= \sum_j  g_{j}\qtya{ a_{j}^\dagger + a_j },
\end{align}
where $a_j$ and $a_j^\dagger$ are the annihilation and creation operators of the $j$-th mode in the bath, 
$\omega_j$ is the frequency of $j$-th mode, 
and $g_j$ is the coupling strength between the system and the $j$-th mode in the bath, which is assumed to be a real number.
Throughout this paper, we use $\hbar = 1$ for simplicity.

The multi-DoF wavefunction $\ket{\Psi(t)}$ is represented by a tensor  
\begin{align}
  \ket{\Psi(t)} = \sum_{i\vec{n}} A_{i n_1 \cdots n_J}(t) \ket{\phi_i} \prod_{j=1}^{J}\ket{\chi^{(j)}_{n_j}},
\end{align}
with tensor elements $A_{i n_1 \cdots n_J}$, where index $i$ runs over the basis of the system while the index $n_j$ runs over the $j$-th harmonic oscillator in the bath.
Here $\ket{\phi_i}$ is the basis of the system while $\ket{\chi^{(j)}_{n_j}}$ is the basis for the $j$-th harmonic oscillator of the bath.
The multi-index $\vec{n} \equiv (n_1, n_2, \ldots, n_J)$,  $n_j = 0, 1, \ldots$ for $j=1, \ldots, J$,  
is used to denote the collection of all indices $n_j$ for the $J$ harmonic oscillators in the bath. 
The full dynamics of the system plus bath are completely described by the time-dependent Schr\"odinger equation (TDSE), 
$\iu \dv{t} \ket{\Psi(t)} = H \ket{\Psi(t)}$.
For the sake of numerical simulation, this approach requires the bath to be discretized by a finite number $J$ of DoFs.
Nevertheless, increasing the DoFs in the bath part increases the cost of explicitly tracking the tensor $A_{i n_1 \cdots n_J}$ exponentially with respect to $J$.

To practically simulate the quantum dynamics of the model,
the ML-MCTDH theory \cite{Wang2003,Wang2021} represents the wavefunction $\ket{\Psi(t)}$ by a tree tensor network with $(J+1)$ DoFs in total.
Further, the TDSE for $\ket{\Psi(t)}$ is decomposed into a set of coupled master equations for each tensor in the tensor network using the time-dependent variational principle.

\subsection{Wavefunction-based methods at finite temperature}\label{sec:ch7-wfn-FT}
To simulate the quantum dynamics of the system in a thermal bath, in general the system needs to be described by a density matrix $\rho(t)$ instead of a wavefunction $\ket{\Psi(t)}$.  
Specifically, if the bath is at thermal equilibrium and the system can initially be described as a pure state, then the initial density matrix of the system plus bath can be written as
\begin{equation}\label{eq:ch7-init-ft}
  \rho(0) = \ket{\psi_\text{S}(0)}\bra{\psi_\text{S}(0)} \otimes \rho_{\text{B}}^{\text{eq}},
\end{equation}
where $\ket{\psi_\text{S}(0)}$ is the initial state of the system, and $\rho_{\text{B}}^{\text{eq}} = Z^{-1} e^{-H_{\text{B}}/(k_{\text{B}}T)}$ is the equilibrium density matrix of the bath.
Here $T$ is the temperature, $k_{\text{B}}$ the Boltzmann constant, and $Z = \Tr (e^{- H_{\text{B}}/(k_{\text{B}}T)})$ the partition function.

Using wavefunction methods, it is still possible to simulate the dynamics of the system in a thermal bosonic bath using purification \cite{Schollwoeck2011,Ren2018}  {or thermofield theory \cite{Takahashi1996, Vega2015, Borrelli2016, Brey2021, Tamascelli2019, Takahashi2024a, Takahashi2025, Harsha2019}}.
In this strategy, the density matrix $\rho(t)$ in the Liouville space is mapped to a wavefunction $\ket{\Psi(t)}$ in an augmented Hilbert space with auxiliary modes.
That is, the original system at temperature $T$ is replaced by a fictitious augmented system at zero temperature, such that the two dynamics coincide.

Specifically, suppose the interaction between the bosonic bath and the system is described by Eq.~\eqref{eq:ch7-sb-h}.
In this case, for a thermal bath at temperature $T$, the effective Hamiltonian for the augmented composite system can be characterized by
\cite{Tamascelli2019,Tokieda2020}
\begin{equation}\label{eq:ch7-heff}
  H_{\text{eff}} = H_{\text{S}} + H_{\text{B}'} + H_{\text{SB}'}
\end{equation}
where 
\begin{align}\label{eq:ch7-heff-term1}
  H_{\text{B}'} &=  \sum_{j} \omega_j  \alpha_{j}^{\dagger} \alpha_{j} - \sum_{j} \omega_j \beta_{j}^{\dagger} \beta_{j},  \\
  H_{\text{SB}'} &= \sum_j  g^+_{j}\qtya{ \alpha_{j}^{\dagger} +  \alpha_j} + \sum_j  g^-_{j}\qtya{ \beta_{j}^{\dagger} + \beta_j},\label{eq:ch7-heff-term2}
\end{align} 
and $\alpha_j$, $\beta_j$ ($\alpha_j^\dagger$, $\beta_j^\dagger$ ) are the bosonic annihilation (creation) operators in the effective bath constructed from the $j$-th modes in the original bath.
Here the effective coupling strength $g_j^+$ and $g_j^-$ are defined as
\begin{equation}\label{eq:ch7-eff-coupling}
  g^{\pm}_{j} = g_{j} \sqrt{\frac{1}{2}\qtya{\coth\qtya{\frac{\omega_j}{2k_\text{B} T}} \pm 1 }}.
\end{equation} 

To see the equivalence between the two approaches, consider the bath correlation function (BCF) $C(t) = \ev{\tilde{X}_\text{B}(t) \tilde{X}_\text{B}(0) \rho_\text{B}^{\text{eq}}}$ which, together with $H_\text{S}$ and $Q_\text{S}$, fully determines the dynamics of the system interacting with a macroscopic Gaussian bath.\cite{Tanimura1990,Ishizaki2009} 
Here $\tilde{O}(t) = \qtya{\mathcal{T} e^{-\iu\int_0^t H_0(t') \dd{t'}}}^\dagger O(t) {\mathcal{T} e^{-\iu\int_0^t H_0(t') \dd{t'}}}$ is the operator in the interaction picture with respect to $H_0(t) = H_\text{S}(t) + H_\text{B}$, and $\mathcal{T}$ is the time-ordering operator.
The main idea is that the effective bath described by Eq.~\eqref{eq:ch7-heff} at zero temperature will have the same BCF as the original bath at temperature $T$.
To see this, for the original bosonic bath, the BCF is determined by \cite{May2011}
\begin{equation}\label{eq:ch7-fdt}
  C(t) = \int_{-\infty}^{+\infty} \mathcal{J}(\omega) (1 + n(\omega, T)) e^{-\iu \omega t}  \dd{\omega}.
\end{equation}
Here $\mathcal{J}(\omega) = \sum_j \qtya{g_j^2 \delta(\omega - \omega_j) - g_j^2 \delta(\omega + \omega_j)}$ 
is the spectral density of the bath, 
and $n(\omega, T) = {1}/({e^{\omega/(\kB T)} - 1})$ is the Bose--Einstein distribution at temperature $T$.
Using the identity $1+n(\omega, T) = \frac{1}{2} \left(\coth\qtya{\frac{\omega}{2\kB T}} + 1\right)$,
we can express $C(t) = \int_{-\infty}^{+\infty} \mathcal{C}(\omega) e^{-\iu \omega t} \dd{\omega}$,
where 
\begin{equation}\label{eq:ch7-spectral-density-eff}
  \mathcal{C}(\omega) = \sum_j \left( (g_j^+)^2 \delta(\omega - \omega_j) + (g_j^-)^2 \delta(\omega + \omega_j)\right)
\end{equation}
is the spectral density of the effective bath.
The advantage of writing it like this is that the initial thermal bath is now mapped to the new bath with all bosonic modes in the vacuum zero-temperature state, and the system is now coupled with the new effective bath at zero temperature with the effective Hamiltonian $H_{\text{SB}'}$.

Notice that the effective Hamiltonian $H_{\text{eff}}$ doubles the DoFs in the original bath.
Suppose that for the original system coupled to a zero temperature bath, the total wavefunction $\ket{\Psi(t)}$ can be represented by a ML-MCTDH ansatz as $\ket{\Psi(t)}$ with $J$ DoFs for the bath.
To capture the influence of a thermal bath, the dynamics of the same system needs to be simulated with a ML-MCTDH ansatz with $2J$ DoFs.
In both cases, the bath is discretized by a finite number of DoFs, where each DoF is explicitly represented by a harmonic oscillator which is Hermitian, and thus, not dissipative.
Therefore, such discretization cannot strictly describe the irreversibility and dissipative nature of an open quantum dynamics for arbitrary long time, but with enough number of DoFs, the discretized model can be used to mimic the dynamics up to a certain time scale.

\subsection{Dissipative density matrix-based method: Hierarchical equations of motion}\label{sec:ch7-dm}
In general, using wave function methods and finite discretization of the bath it is not possible to fully describe the open quantum dynamics even with purification strategies.
This type of physical processes are more naturally described in terms of master equations satisfied by the reduced density matrix of the system $\rho_\text{S}(t)$, which include dissipative terms that do not conserve the unitary evolution of the system.
The hierarchical equations of motion (HEOM) is one of the most advanced and successful approach for simulating the open quantum system coupled to a macroscopic thermal dissipative bath in a numerically exact manner.\cite{Tanimura1990,Tanimura2020} 
In the HEOM, the BCF is decomposed into a series of complex exponentials as  
\begin{equation}\label{eq:ch7-bcf}
  C(t) = \sum_{k=1}^K c_k e^{\gamma_k t}, \qand C^{\star}(t) = \sum_{k=1}^K \bar{c}_k e^{\gamma_k t},
\end{equation}
where $c_k$, $\bar{c}_k$ and $\gamma_k$ are complex numbers.
This decomposition is then used to map the open quantum dynamics into the dynamics of $\rho_\text{S}(t)$ and $K$ bexcitons, which are bosonic fictitious quasiparticles that oscillate and decay to exactly capture the influence of the bath. 
Here we briefly summarize the bexcitonic extended density operator $\ket{\varrho(t)}$ and its dynamics described by the HEOM, as detailed in Ref.~\onlinecite{Chen2024}.

Suppose that at the initial time the system is in a separable state $\rho(0) = \rho_{\text{S}}(0) \otimes \rho_{\text{B}}^{\text{eq}}$, where $\rho_{\text{S}}(0)$ is the initial reduced density matrix of the system.
We define the bexcitonic extended density operator $\ket{{\varrho}(t)}$, which is a vector of density operators.
In $\ket{{\varrho}(t)}$, each density operator is of the same dimension as the system's reduced density operator $\rho_{\text{S}}(t)$.
$\ket{{\varrho}(t)} = \sum_{n_1 n_2 \cdots n_K} \rho_{n_1 n_2 \cdots n_K}(t) \ket{n_1 n_2 \cdots n_K}$
is defined on a $K$-boson basis $\set{\ket{n_1 n_2 \cdots n_K}}$ with $n_k = 0,\ 1,\ \ldots$ for $k=1,\ \ldots,\ K$, where $\ket{n_k}$ is the $n_k$-th Fock state of the $k$-th auxiliary bosonic bexciton, and each $\rho_{n_1 n_2 \cdots n_K}(t)$ is an auxiliary density matrix.

The bexcitonic generalization of the HEOM describes the dynamics of $\ket{{\varrho}(t)}$ as\cite{Chen2024}
\begin{equation} \label{eq:ch7-heom}
  \begin{aligned}
    \dv{t} \ket{\varrho(t)} &= \qtya{ -\iu H_\text{S}^{\times}(t) + \sum_{k=1}^K \mathcal{D}_{k} }\ket{\varrho(t)},
  \end{aligned}
\end{equation}
with 
\begin{equation}
  \begin{multlined}  
    \mathcal{D}_{k} =
    \gamma_{k} {\alpha}^\dagger_k \hat{\alpha}_{k} + 
    Q_\text{S}^{>} \qtya{c_k {z}_{k}^{-1} \hat{\alpha}^\dagger_k - \hat{\alpha}_k {z}_{k}} \\
    - Q_\text{S}^{<} \qtya{\bar{c}_k {z}_{k}^{-1} \hat{\alpha}^\dagger_k - \hat{\alpha}_k {z}_{k}},
  \end{multlined}
  \end{equation}
and the initial condition
\begin{equation}
  \ket{\varrho(0)} = \rho_{\text{S}}(0) \otimes \ket{\underbrace{0\cdots 0}_K}.
\end{equation}
Here we have adopted the bosonic annihilation operator $\hat{\alpha}_k$ and creation operator $\hat{\alpha}_k^\dagger$ to represent the $k$-th auxiliary bosonic bexciton introduced in the HEOM,
and $z_k$ is a non-zero complex number.
We have also used $A^> B = A B$ and $A^< B = B A $ for the left and right multiplication.
To obtain the reduced density matrix of the system, one needs to project $\ket{\varrho(t)}$ onto the basis vector $\ket{0  \cdots 0}$ as
\begin{equation}
  \rho_{\text{S}}(t) = \ip{0 \cdots 0}{{\varrho}(t)}.
\end{equation}

The HEOM thus introduces $K$ additional DoFs to track the history of the system-bath interaction.
Since $K$ is determined by the number of terms in the BCF decomposition, each of which can characterize the dissipative nature of the bath,  it is possible to use a small number of auxiliary DoFs to capture the dynamics of the system.
Similar to the wavefunction-based methods, the computational cost of the HEOM also increases exponential with respect to $K$. 
As in ML-MCTDH,  we can develop a tensor network decomposition of the HEOM to reduce its computational cost.\cite{Shi2018, Borrelli2019, Borrelli2021, Yan2021, Li2022, Chen2025} 
In this approach, the $(K+2)$-order tensor $\ket{{\varrho}(t)}$ is represented by a tree tensor network, and the dynamics of the system is decomposed by a set of master equations for each tensor in the tensor network, as detailed in Ref.~\onlinecite{Chen2025}.

\section{Discretization strategies for a dissipative bath}\label{sec:ch7-strategy}
In this section, we discuss discretization strategies employed in ML-MCTDH and TTN-HEOM to capture the influence of the dissipative bath.

\subsection{Discretization of the bath for the explicit wavefunction-based method}
To use the wavefunction-based simulation method to simulate the quantum dynamics of a system interacting with a thermal bath with a continuous spectral density $\mathcal{J}(\omega)$ using purification,
the bath needs to be discretized by a set of harmonic oscillators with both positive and negative frequencies in a bounded range $(-\omega_\text{C}, \omega_\text{C})$, as the effective spectral density $\mathcal{C}(\omega)$ in Eq.~\eqref{eq:ch7-spectral-density-eff} defines on both positive and negative frequencies.
Suppose the discretized frequencies $\omega_k$, $k = \pm 1,\ \pm 2,\ \ldots,\ \pm K$ are sorted as $\omega_{-K} < \cdots < \omega_{-1} < 0 < \omega_1 < \cdots < \omega_K$,  and the number of DoFs and the absolute value of the cutoff frequency $\omega_\text{C}$ for both positive and negative frequencies are the same.
The effective coupling strength of each mode with the thermal effect included is determined by the bath spectral density near those frequencies. That is, in Eq.~\eqref{eq:ch7-fdt}, we use the approximation
\begin{equation}
\mathcal{C}(\omega)\dd{\omega} = \mathcal{J}(\omega) (1 + n(\omega, T)) \dd{\omega}
\approx
\sum_{k=\pm 1}^{\pm K}  g_k^2 \delta(\omega - \omega_k) \Delta_k,
\end{equation}
where
\begin{align}
  \label{eq:ch7-g-discrete}
  g_k &= \sqrt{\mathcal{J}(\omega_k) (1 + n(\omega_k, T)) \Delta_k}, \quad\text{and} \\
  \label{eq:ch7-delta}
  \Delta_k &= \abs{\omega_{k+1} - \omega_{k-1}}/2.
\end{align}
We also let $\omega_{0} = 0$ and $\omega_{\pm (K+1)} = \omega_\text{C}$ at the boundaries $k = \pm 1$ and $\pm K$ such that the formula of $\Delta_k$ holds for all $k$.
In this way, the effective thermal dynamics of the system can be described by the Hamiltonian with $2K$ bath DoFs.  The effective Hamiltonian is
\begin{equation}
  H_{\text{eff}} = H_{\text{S}} + H_{\text{B}'} + H_{\text{SB}'},
\end{equation}
where
\begin{align}\label{eq:ch7-heff-db}
  H_{\text{B}'} &= \sum_{k=1}^{K} \qtya{ \omega_k a_{k}^\dagger a_{k} +  \omega_{-k} a_{-k}^{\dagger} a_{-k}},\qand\\
  \label{eq:ch7-hef-dsb}
  H_{\text{SB}'} &= Q_\text{S} \sum_{k=1}^{K} \qtya{g_k (a_k^\dagger + a_k) + {g}_{-k} (a_{-k}^{\dagger} + a_{-k})},
\end{align} 
which has the form of the effective Hamiltonian in Eq.~\eqref{eq:ch7-heff}, but may have different discretization frequencies $\omega_k$ and $\omega_{-k}$ for the positive and negative domain.

To obtain a set of discretized frequencies $\omega_k$, one intuitive way is to first choose a set of discretized frequencies for $\omega_k>0$ in a logarithmic space on $(\omega_\text{min}, \omega_\text{max})$, as $\log \omega_k = \log \omega_\text{min} + (k-1) \Delta $ with $\Delta  = (\log \omega_\text{max} - \log \omega_\text{min})/K$.
For the negative frequency, we choose $\omega_{-k} = -\omega_k$ for $k = 1,\ \ldots,\ K$.
This is denoted as the \emph{logarithmic discretization} in the rest of this paper.

Another strategy is to balance the area under the effective spectral density $\mathcal{J}(\omega) (1 + n(\omega, T))$ for each discretized frequency $\omega_k$.
That is, for each $\omega_k$ with $k = 1,\ \ldots,\ K$, we first calculate the overall area under the effective spectral density $\mathcal{J}(\omega) (1 + n(\omega, T))$ as
\begin{equation}
  A = \int_{0}^{\omega_\text{C}} \mathcal{J}(\omega) (1 + n(\omega, T)) \dd{\omega},
\end{equation}
and scan over the frequency range $(0, \omega_\text{C})$ to find the discretized frequencies $\omega_k$ such that
\begin{equation}\label{eq:ch7-equalized-plus}
  \int_{0}^{\omega_k} \mathcal{J}(\omega) (1 + n(\omega, T)) \dd{\omega} = \frac{k+1/2}{K} A,\quad k = 1,\ \ldots,\ K.
\end{equation}
with $g_k = \sqrt{A/K}$ as the effective coupling strength.
In this way, the area between $\omega_{k-1}$ and $\omega_k$ is equalized, \emph{i.e.},
\begin{equation}\label{eq:ch7-equalized}
  \int_{\omega_{k-1}}^{\omega_k} \mathcal{J}(\omega) (1 + n(\omega, T)) \dd{\omega} = \frac{A}{K},\quad k = 1,\ \ldots,\ K.
\end{equation}
For the negative frequencies, $k = -1,\ \ldots,\ -K$, we use the similar discretization strategy but with $A' = \int_{-\omega_\text{C}}^{0} \mathcal{J}(\omega) (1 + n(\omega, T)) \dd{\omega}$, and hence, the discretized frequencies $\omega_k$ satisfy
\begin{equation}\label{eq:ch7-equalized-minus}
  \begin{multlined}
    \int_{\omega_k}^{0} \mathcal{J}(\omega) (1 + n(\omega, T)) \dd{\omega} = \frac{k+1/2}{K} A',\\ k = -1,\ \ldots,\ -K,
  \end{multlined}
\end{equation}
and use $g_k = \sqrt{A'/K}$ as the effective coupling strength.

The criteria in Eqs.~\eqref{eq:ch7-equalized-plus} and \eqref{eq:ch7-equalized-minus} is similar to the one used by Makri and coworkers for discretizing the bath spectral density,\cite{Walters2017}  but it also includes the thermal effect from $n(\omega, T)$.
Notice that at the high temperature limit, the spectral density $\mathcal{J}(\omega) (1 + n(\omega, T))$ is proportional to $(\mathcal{J}(\omega)/\omega + \frac{1}{2})$, which corresponds to the case where the reorganization energy is equalized for each discretized frequency $\omega_k$.
For the rest of this paper, we will call this discretization strategy the \emph{equalized discretization}.

 {%
A third strategy to obtain a discretization of the bath is the \emph{bath spectral density orthogonal} (BSDO) \cite{Vega2015} method, which can provide an accurate BCF decomposition with a small number of discretized modes. In this method, a discrete representation of the BCF is constructed by Gauss quadrature using the spectral density as the weight of a quadrature, which can significantly improve frequency sampling as it enables to construct an efficient set of polynomial interpolants. The BSDO can also be considered as the star-mapping of the effective chain model of the finite temperature bath used in the thermalized time-evolving density operator with orthogonal polynomials (T-TEDOPA) method \cite{Chin2010a, Tamascelli2019}. This method is discussed in details in Refs.~\onlinecite{Vega2015, Takahashi2025, Chin2010a, Tamascelli2019}. In particular, we employ the strategy detailed in Refs.~\onlinecite{Chin2010a, Tamascelli2019} followed by a diagonalization to transform it from the chain map to the star map of the harmonic environment.}

To evaluate the performance of the discretization strategy, we compare the discretized effective spectral density, $\bar{\mathcal{C}}(\omega)$, calculated from the bath correlation function of the discretized model to the exact $\mathcal{C}(\omega)$. Here the $\bar{\mathcal{C}}(\omega)$ is evaluated from the Fourier transform 
\begin{equation}
  \bar{\mathcal{C}}(\omega) = \frac{1}{2\pi}\int_{-\infty}^{+\infty}  \bar{C}(t) e^{\iu \omega t} \dd{t},
\end{equation}
of the discretized bath correlation function
\begin{equation}
  \bar{C}(t) =\sum_{j} {g}_j^2 e^{-\iu {\omega}_j t}.
\end{equation}
Here ${g}_j$ is the effective coupling strength of the discretized model, ${\omega}_j$ is the discretized frequency, and $j = \pm k$ for $k = 1,\ \ldots,\ K$. Notice that the overall number of discretized mode $J = 2K$ as we have both positive and negative frequencies.

\subsection{Decomposition of the bath correlation function as an implicit discretization}
As noted in Sec.~\ref{sec:ch7-dm}, in the HEOM the discretization of the bath described by a continuous spectral density $\mathcal{J}(\omega)$ consists of a BCF $C(t)$ with a few exponentially decaying terms in the BCF decomposition Eq.~\eqref{eq:ch7-bcf}.
To characterize such a macroscopic bath, a commonly used bath spectral density is the Drude--Lorentz (DL) bath, which is often used to model the solvent modes.
The DL model \cite{Caldeira1981, Grabert1988, Tanimura1990}
\begin{equation}\label{eq:ch7-dl}
  \mathcal{J}_\text{D}(\omega) = \frac{2\lambda_\text{D}}{\pi} \frac{\gamma_\text{D}\omega}{\omega^2 + \gamma_\text{D}^2}
\end{equation}
models an Ohmic bath with reorganization energy $\lambda_\text{D}$ and characteristic frequency $\gamma_\text{D}$.
Other models include the Brownian oscillator  {which is} used to describe possibly damped discrete vibrational baths.\cite{Liu2014, Dunn2019, Yan2020, Krug2023}  {In this case, the spectral density is 
\begin{equation}
     J_\mathrm{B}(\omega)= \frac{4\lambda_{\text{B}}}{\pi} \frac{\gamma_{\text{B}} \Omega^2 \omega}{(\omega^2 - \Omega^2)^2 + 4\gamma_{\text{B}}^2 \omega^2},
\end{equation}
with $\omega_1 \equiv \sqrt{\Omega^2 - \gamma_{\text{B}}^{2}} > 0$ the effective oscillation frequency of the bath mode, $\lambda_\text{B}$ the reorganization energy, and $\gamma_\text{B}$ the damping rate.
}

To obtain the decomposition of BCF, we evaluate Eq.~\eqref{eq:ch7-fdt} using the residue theorem through analytical continuation and expanding $(1 + n(\omega, T))$ through a decomposition using rational functions.\cite{Hu2010, Zheng2009, Cui2019,Zhang2020,Xu2022}  
\begin{equation}\label{eq:ch7-cexpansion}
  \begin{aligned}
  C(t) &= - 2 \pi \iu \sum_{i} {\mathop {\operatorname {Res}}_{z=\zeta_i}}[\mathcal{J}(z)] (1 + n(\zeta_i, T)) e^{-\iu \zeta_i t}\\
  &\quad- 2 \pi \iu \sum_{j} {\mathop {\operatorname {Res}}_{z=\xi_j}}[1 + n(z, T)] \mathcal{J}({\xi_j}) e^{-\iu  \xi_j t},
  \end{aligned}
\end{equation}
where $\{\zeta_i\}$ are the first order poles of $\mathcal{J}(\omega)$ and  $\{\xi_j\}$ those of $(1 + n(\omega, T))$ in the lower-half complex plane. 
These expansions satisfy Eq.~\eqref{eq:ch7-bcf} with each term defining a feature.  
Eq.~\eqref{eq:ch7-cexpansion} leads to overall exponentially decaying BCF as the poles satisfy $\Im \zeta_i$ and $\Im  \xi_j < 0$.

 {As an example}, for the DL bath, the poles from $\mathcal{J}(\omega)$ is $\zeta_i = -\iu \gamma_\text{D}$, which corresponds to one term in Eq.~\eqref{eq:ch7-cexpansion} as $c_1 e^{- \gamma_\text{D} t}$ where $c_1 = \lambda \gamma_\text{D} \qtya{\cot(\gamma_\text{D}/(2\kB T))-\iu}$.
This term decays exponentially on a time scale $\tau = \gamma_\text{D}^{-1}$, which corresponds to the high-temperature limit of the bath and it is inherently dissipative. 
On the other hand, the poles from $(1 + n(\omega, T))$ are of infinite order, but can be approximated by a series of rational functions with first order poles only. 
For instance, the poles $\xi_j$ can be approximated as $\xi_j = -\iu \omega_j$ with $\omega_j = 2\pi j \kB T >0$, $j = 1,\ \ldots$, if one adopts the Matsubara series.\cite{Zheng2009} 
Other rational functions series are also possible, such as the Pad\'e approximant,\cite{Hu2010}  which also gives the poles $\xi_j = -\iu \omega_j$ with $\omega_j >0$.
One can also directly use a series of rational functions to approximate the effective spectral density $\mathcal{J}(\omega) (1 + n(\omega, T))$ with first-order poles only to evaluate Eq.~\eqref{eq:ch7-fdt}, as suggested in recent literature.\cite{Xu2022, LeDe2024} 
Alternatively, directly fitting the bath correlation functions with a sum of complex exponential functions is also possible.\cite{Takahashi2024} 

{%
We also note that the Drude--Lorentz spectral density exhibits a singularity and an unphysical response function at $t=0$.\cite{Ishizaki2020} While it is primarily intended to describe long-time dynamics, this feature can make it challenging for explicit discretization schemes to accurately reproduce the short-time dynamics with a small number of discretization modes.
}

We use an implicit discretization strategy based on the Pad\'e $[(N-1)/N]$ approximant as the rational function series for $(1 + n(\omega, T))$ \cite{Hu2010} to compare the results from the explicit discretization.
For a DL bath, this gives $N+1$ terms in the BCF decomposition Eq.~\eqref{eq:ch7-bcf} for HEOM, where the first term is from the pole from $\mathcal{J}(\omega)$, and the other $N$ terms are from the poles from $(1 + n(\omega, T))$.

\section{Results}\label{sec:ch7-results}

\subsection{ {Drude--Lorentz environment}}
\label{sec:ch7-dl}

 {We first focus on the implicit and explicit discretization of the Drude--Lorentz model, which is a widely used model of thermal environments. Section~\ref{sec:ch7-br} discusses the discretization of the Brownian oscillator model.} 

\subsubsection{ {Discretization of the spectral density}}

\begin{figure}[tbp]
  \centering
  \includegraphics[width=\linewidth]{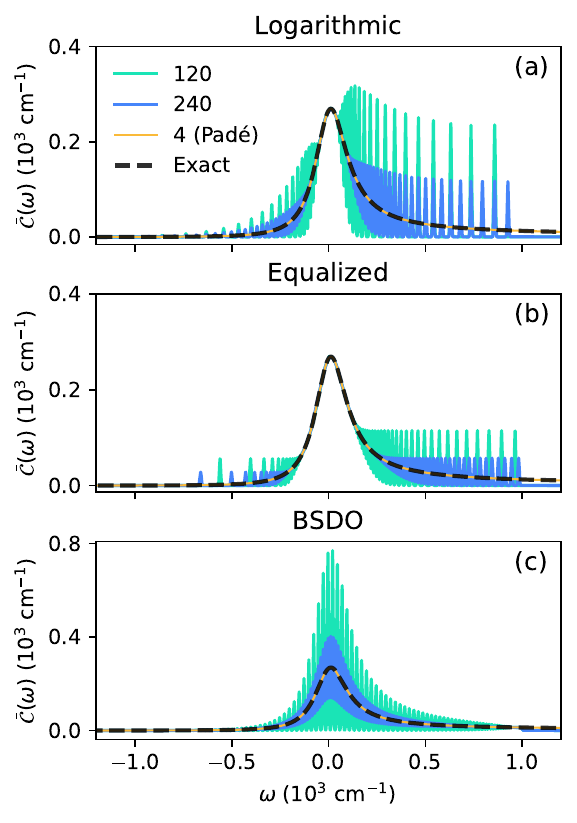}
  \caption{Discretized spectral density  {developed from the }  {(a) logarithmic, (b) equalized, and (c) BSDO discretization strategies}.
  The spectral density $\mathcal{J}(\omega)$ is the Drude--Lorentz (DL) one with reorganization energy $\lambda_\text{D} = 200~\invcm$ and characteristic frequency $\gamma_\text{D} = 100~\invcm$, and temperature is set to be $T = 300~\text{K}$.
  The cutoff frequency of the discretization is set at $1000~\invcm$.
  For the logarithmic discretization, the minimum frequency $\omega_\text{min}$ is set to be $0.01~\invcm$.
  The number of discretized DoFs is set to be  {$J=120,\ 240$} for each panel.
  The implicit discretization using the Pad\'e $[2/3]$ approximant with $K=4$ (in yellow solid) and the exact target $\mathcal{J}(\omega)(1+n(\omega, T))$ (in black dashed) are also plotted for comparison.}
  \label{fig:sd-dof}
\end{figure}

Figure~\ref{fig:sd-dof} shows the discretized spectral density $\bar{\mathcal{C}}(\omega)$ calculated from the bath correlation function of the discretized model with the logarithmic, equalized {, and BSDO} discretization {s}. 
For  {all} methods, the cutoff frequency of the discretization is set at $1000~\invcm$. For the logarithmic discretization, the minimum frequency $\omega_\text{min}$ is set to be $0.01~\invcm$.
To numerically evaluate the effective spectral density $\bar{\mathcal{C}}(\omega)$ from the BCF that oscillates instead of decaying in the time domain, we apply a Gaussian window function ${G}(t) = \exp(-{t^2}/({2\sigma^2}))$ with a width $\sigma$ as
\begin{equation}
  \bar{\mathcal{C}}(\omega) = \frac{1}{2\pi}\int_{-\infty}^{+\infty} {G}(t) \bar{C}(t) e^{\iu \omega t} \dd{t}.
\end{equation}
We choose $\sigma = 300~\text{fs}$ for the simulation, which results in a broadening of the spectral density $\bar{\mathcal{C}}(\omega)$ with a width of $\sigma^{-1} \approx 3~\invcm$.
 {Figure}~\ref{fig:sd-dof} shows the discretized spectral density $\bar{\mathcal{C}}(\omega)$ with such broadening using  {(a) logarithmic, (b) equalized, and (c) BSDO discretization} with different number of discretized DoFs  {$J=120,\ 240$}.

The figure also show the converged implicit discretization with Pad\'e $[2/3]$ approximant of $(1+n(\omega, T))$ in each panel for comparison.
This corresponds to the $K=4$ case in HEOM.
As shown,  {for the logarithmic and equalized discretization strategies, } the exact effective spectral density $\mathcal{J}(\omega)(1+n(\omega, T))$ can be well-approximated by the discretized spectral density $\bar{\mathcal{C}}(\omega)$ for the frequencies near zero as the number of discretized DoFs increases.
 {In turn, the BSDO discretization recovers more of the high frequency part before the cutoff frequency.}
However, for the frequencies far away from zero, the discretized spectral density $\bar{\mathcal{C}}(\omega)$ deviates from the exact one even when a large number of discretized DoFs are used.

 {%
Figures S1--S3(a, b) shows the BCF decomposition in the time domain using the equalized, logarithmic, and BSDO discretization strategies.
The equalized and logarithmic discretization strategy can provide the overall trend of the BCF decay at the short time for the Drude--Lorentz environment. 
For the longer time, they gradually deviate from the exact BCF as the number of discretized modes is limited. 
By contrast, the BSDO discretization strategy can provide a more accurate BCF decomposition with the same number of discretized modes, especially for the short-time regime. However, when it deviates from the exact BCF at longer times, it will deviate more significantly than the equalized discretization strategy. 
}

\subsubsection{Quantum dynamics of a two-level system}\label{sec:ch7-dl-tls}
Consider first the dynamics of a two-level system coupled to a DL bath with the spectral density in Eq.~\eqref{eq:ch7-dl}.
The two-level system $\{\ket{0}, \ket{1}\}$ is described by the Hamiltonian 
\begin{equation}
  H_\text{S} = \frac{E}{2} \sigma_z + V \sigma_x,
\end{equation}
where $\sigma_z$ and $\sigma_x$ are the Pauli matrices, $E$ is the energy difference between the two states, and $V$ is the coupling strength between the two states.
The system operator $Q_\text{S}$ in the system-bath interaction Hamiltonian is chosen as $Q_\text{S} = \sigma_z/2$.
Notice that when $V = 0$, the system is in the pure dephasing limit, and when $E = 0$, the system is in a dissipative relaxation limit.
We will first discuss the pure dephasing dynamics and then the relaxation dynamics.

\paragraph{Pure dephasing limit.}\label{sec:ch7-sbm-model}
In the pure dephasing limit, 
the population of the energy eigenstates is fixed during the dynamics as $\comm{H_\text{S}}{H_\text{SB}} = 0$.
Suppose that the system is initially prepared in a superposition  $\ket{\psi_\text{S}(0)} = (\ket{0} + \ket{1})/\sqrt{2}$.
In this case, the dynamics of the coherence $\abs{[\rho_{\text{S}}]_{01}}$ can be exactly solved as \cite{Schlosshauer2007}
\begin{equation}
  \begin{multlined}
    \abs{[\rho_{\text{S}}]_{01}} = \\
    \exp\qtya{ - \int_0^{+\infty} J(\omega)\coth\qtya{\frac{\omega}{2\kB T}}\frac{1- \cos(\omega t)}{\omega^2} \dd{\omega}}.
  \end{multlined}
\end{equation}
Therefore, the dynamics of the coherence $\abs{[\rho_{\text{S}}]_{01}}$ and population ${[\rho_{\text{S}}]_{00}}$ can be used as a benchmark for the simulation of the two-level system for different simulation methods.

The two-level system is coupled to a DL bath with the spectral density $\mathcal{J}(\omega)$ as shown in Fig.~\ref{fig:sd-dof}.
We use the ML-MCTDH with logarithmic, equalized and  {BSDO} discretization, as well as the TTN-HEOM with the Pad\'e approximant, to simulate the dynamics of such a two-level system using the \texttt{TENSO} package.\cite{Chen2025} 
In the simulations, a balanced tree structure with maximal bond threshold $R=64$ was used for both TTN-HEOM and ML-MCTDH with a mixed propagation scheme that combines both two-site projector splitting propagation algorithm and direct integration of the master equations from the time-dependent variational principle.
The dimension of the truncated Fock basis $\{\ket{n_k}\}_{n_k=0}^{N-1}$ of each auxiliary mode in the HEOM is set with $N=20$, while for each discretized DoF in the ML-MCTDH the dimension  {(the dimension of each primitive basis)} is set to be $N=2$.

\begin{figure}[tbp]
  \centering
  \includegraphics[width=\linewidth]{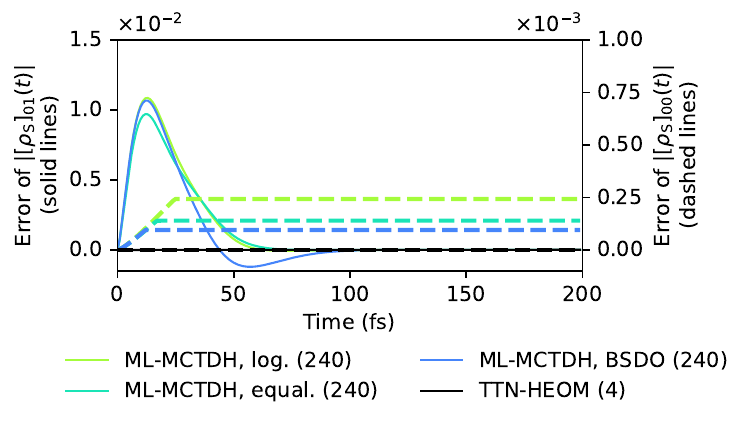}
  \caption{Errors in capturing the pure dephasing dynamics of a two-level system using TTN-HEOM and ML-MCTDH. 
  The two-level system is prepared in a superposition of the two states $\ket{\psi(0)} = (\ket{0} + \ket{1})/\sqrt{2}$ and coupled to a DL bath at $300$~K. 
  Here the reorganization energy is $\lambda_\text{D} = 200~\invcm$ and the characteristic frequency $\gamma_\text{D} = 100~\invcm$.
  ML-MCTDH is used to simulate the discretized dynamics with  {logarithmic (log.), equalized (equal.), and BSDO} discretization strategies with $J=240$ and cutoff frequency $1000~\invcm$.
  The TTN-HEOM is used to simulate the implicit discretization using the Pad\'e $[2/3]$ approximant with $K=4$.
  The solid lines are the error of the coherence $\abs{[\rho_{\text{S}}]_{01}(t)}$, while the dashed lines are the error of the population $\abs{[\rho_{\text{S}}]_{00}(t)}$. 
  The number in the bracket for each label indicates the number of discretized DoFs used in the simulation.
  }
  \label{fig:toy1}
\end{figure}

The performance of TTN-HEOM and ML-MCTDH in this pure dephasing dynamics is shown in Fig.~\ref{fig:toy1}.
In the figure, the ML-MCTDH results from the logarithmic, equalized {, and BSDO} discretizations with $J=240$ DoFs and cutoff frequency $1000~\invcm$ are shown in blue, cyan {, and green} lines, respectively.
The TTN-HEOM results from Pad\'e $[2/3]$ approximant with $K=4$ are in black.
The comparison to the analytical solution shows that both the explicit and implicit discretization can capture the pure dephasing dynamics of the two-level system with an error of less than $1\%$.
However, the explicit discretization strategy is more difficult to converge to the exact solution even with a larger number of discretized DoFs and overall Hilbert space dimension compared to the implicit discretization strategy.
Further, when using the explicit discretization strategy, the logarithmic, equalized  {and BSDO} discretizations are giving similar results when the number of discretized DoFs is large enough.
This is expected as the  {three} discretization strategies are equivalent in the limit of large number of discretized DoFs.

\paragraph{Relaxation dynamics.}
\label{sec:ch7-sbm-relaxation}
We now focus on the relaxation case with $V =1000~\invcm$ and $E = 0$.
We set the initial state of the system to be $\ket{\psi_\text{S}(0)} = (\ket{0} + \ket{1})/\sqrt{2}$.
In this case, the system is subject to both the dephasing and relaxation effects due to the system-bath interaction.
Due to the symmetry of the system, the population $[\rho_\text{S}]_{00}(t)$ and $[\rho_\text{S}]_{11}(t)$ remain fixed at $1/2$ during the dynamics even during the thermalization process.
On the other hand, for the system Hamiltonian $H_\text{S}$, its two energy eigenstates $\ket{e}$ and $\ket{g}$ are
\begin{equation}
  \ket{g} = \frac{1}{\sqrt{2}}(\ket{0} - \ket{1}),
  \quad 
  \ket{e} = \frac{1}{\sqrt{2}}(\ket{0} + \ket{1}).
\end{equation}
In this case, the system is initially prepared in the excited state $\ket{\psi_\text{S}(0)} = \ket{e}$, and the thermal state $\rho_\text{S}^\text{eq} =
e^{-H_\text{S}/\kB T}/\Tr e^{-H_\text{S}/\kB T}$ is close to $\ket{g}$ 
as  $V/(\kB T) \approx 4.80$ and $[\rho_\text{S}^\text{eq}]_{gg} \approx 0.99993$.
This means that the system is expected to relax to the ground state $\ket{g}$ with a very small population remaining in the initial excited state $\ket{e}$.

\begin{figure}[tbp]
  \centering
  \includegraphics[width=\linewidth]{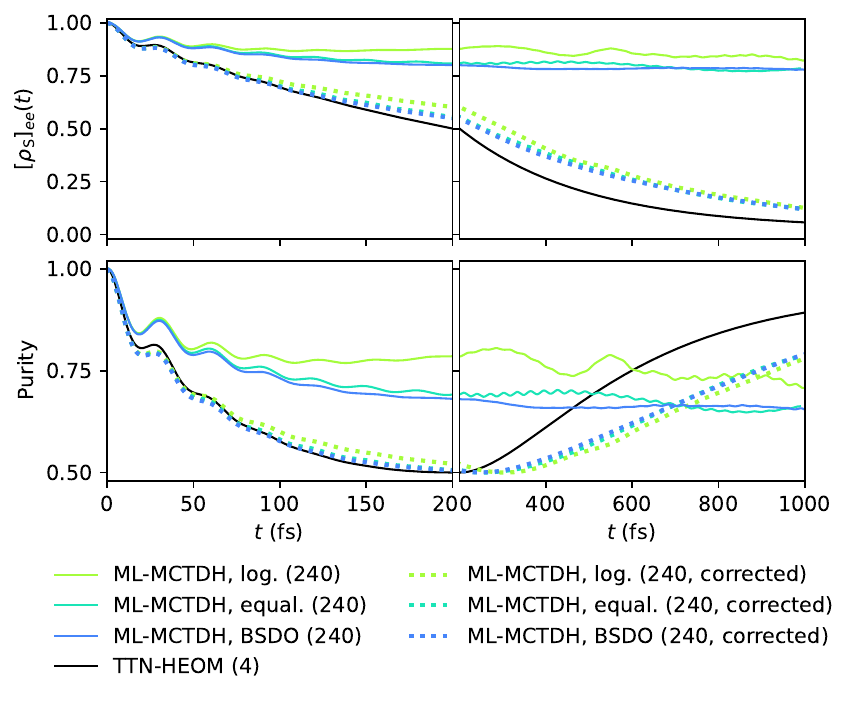}
  \caption{Dissipative dynamics of a two-level system with relaxation using different methods. The two-level system is prepared in $\ket{\psi(0)} = \ket{e} = (\ket{0} + \ket{1})/\sqrt{2}$ and is coupled with a DL bath at $300$~K. Here the reorganization energy $\lambda_\text{D} = 200~\invcm$ and the {characteristic} frequency $\gamma_\text{D} = 100~\invcm$.
  ML-MCTDH are used to simulate the discretized dynamics with  {logarithmic, equalized, and BSDO} discretization strategies with $J=240$ (solid lines) and cutoff frequency $1000~\invcm$.
  The TTN-HEOM is used to simulate the implicit discretization using the Pad\'e $[2/3]$ approximant with $K=4$.
  The upper panels show the population in the excited state  {$[\rho_\text{S}]_{ee}(t)$} of $H_\text{S}$, and the lower panels show the coherence purity $\Tr\rho^2_\text{S}(t)$.
  The dotted lines are the results from the ML-MCTDH with logarithmic, equalized  {and BSDO} discretization strategies with $J=240$ and the correction Eq.~\eqref{eq:ch7-eta} applied.
  The number in the bracket for each label indicates the number of discretized DoFs used in the simulation.
  }
  \label{fig:toy2}
\end{figure}

Similar to the pure dephasing case, we use the ML-MCTDH with  logarithmic, equalized   {and BSDO} discretization strategies with $J=240$ discretized DoFs and cutoff frequency $1000~\invcm$, 
as well as the HEOM with the Pad\'e $[2/3]$ approximant to simulate the dynamics of such a two-level system. 
The dimension of the basis for each auxiliary mode in the HEOM is set to be  {$20$}, while for each discretized DoF in the ML-MCTDH is set to be $2$.
The dynamics of population  {$[\rho_\text{S}]_{ee}(t)$} and purity $\Tr\rho^2_\text{S}(t)$, used to characterize the relaxation dynamics of the two-level system, are shown in Fig.~\ref{fig:toy2}.
The population in the excited state  {$[\rho_\text{S}]_{ee}(t)$} of $H_\text{S}$ is shown in the upper panels, and the coherence purity $\Tr\rho^2_\text{S}(t)$ in the lower panels.

The TTN-HEOM can capture the relaxation dynamics even near the  state with minimum purity with only $4$ auxiliary modes, while the ML-MCTDH with explicit discretization cannot capture the relaxation dynamics even with  {$240$} discretized DoFs.
This may be due to the differences in the tails of the effective spectral density ${\mathcal{C}}(\omega)$, as the relaxation dynamics is determined by the spectral density $\mathcal{J}(\Omega)$ evaluated at the resonance frequency $\Omega=2V$ dictated by the energy difference between the two energy eigenstates of the system Hamiltonian $H_\text{S}$.\cite{Korol2025} 

To correct the dynamics, the rate correction for the excited state population is 
\begin{equation}
  \frac{[\rho^\text{c}_{\text{S}}(t)]_{ee} - [\rho^{\text{eq}}_{\text{S}}]_{ee}}{[\rho_{\text{S}}(t)]_{ee} - [\rho^{\text{eq}}_{\text{S}}]_{ee}} \approx\frac{[\rho^\text{c}_{\text{S}}(t)]_{ee}}{[\rho_{\text{S}}(t)]_{ee}} = e^{- \eta t},
\end{equation}
where $\rho^\text{c}_{\text{S}}(t)$ is the corrected density matrix, and
\begin{equation}\label{eq:ch7-eta}
  \eta = \frac{\pi}{2} \mathcal{J}(\Omega) \coth\qtya{\frac{\Omega}{2\kB T}}.
\end{equation}
Therefore, the corrected $[\rho^\text{c}_{\text{S}}(t)]_{ee} = p_e(t)[\rho_{\text{S}}(t)]_{ee}$ with $p_e(t) = e^{-\eta t}$.
For other matrix elements in $\rho^\text{c}_{\text{S}}(t)$, the ground state population is obtained from $[\rho^\text{c}_{\text{S}}(t)]_{gg} = 1- [\rho^\text{c}_{\text{S}}(t)]_{ee}$,
and the correction factor $p_g(t) = [\rho^\text{c}_{\text{S}}(t)]_{gg}/[\rho_{\text{S}}(t)]_{gg}$.
The corrected $\abs{[\rho^\text{c}_{\text{S}}(t)]_{ge}}$ and $\abs{[\rho^\text{c}_{\text{S}}(t)]_{ge}}$ are assumed to be 
$\abs{[\rho^\text{c}_{\text{S}}(t)]_{ge}} = \sqrt{p_g(t) p_e(t)} \abs{[\rho_{\text{S}}(t)]_{ge}}$ and $\abs{[\rho^\text{c}_{\text{S}}(t)]_{eg}} = \sqrt{p_g(t) p_e(t)} \abs{[\rho_{\text{S}}(t)]_{eg}}$.
The results of the ML-MCTDH with the logarithmic, equalized  {and BSDO} discretization strategies with $J=240$ and this correction applied are shown in Fig.~\ref{fig:toy2} in dotted blue, cyan  {and green} lines, respectively.
As shown in the figure, with such correction, the ML-MCTDH simulations using  logarithmic, equalized  {and BSDO} discretization strategies capture the relaxation dynamics for the first 100~fs, 
and the correction makes the system relax towards the thermal state $\rho_\text{S}^\text{eq}$.
However, the relaxation rate of the excited state population  {$[\rho_\text{S}]_{ee}(t)$} is slower than the one from the TTN-HEOM, suggesting neglected influence from the tail of the spectral density beyond Eq.~\eqref{eq:ch7-eta}.

 {Note that the ML-MCTDH results can be computed with a small primitive basis size (2), which already offers converged results. The reason is that when the number of discretized modes is large, the coupling strength to each individual mode is small, and so is the excitation of each bath mode.
To demonstrate this explicitly, in Fig.~S4 we show the numerical convergence check for the relaxation case. As shown, with a discretization with 120 modes, the primitive basis size of 2 for each mode is sufficient to achieve convergence with larger primitive basis size within the first 400 fs of dynamics, yet still deviates from the TTN-HEOM results.}

\begin{figure}
  \centering
  \includegraphics[width=\linewidth]{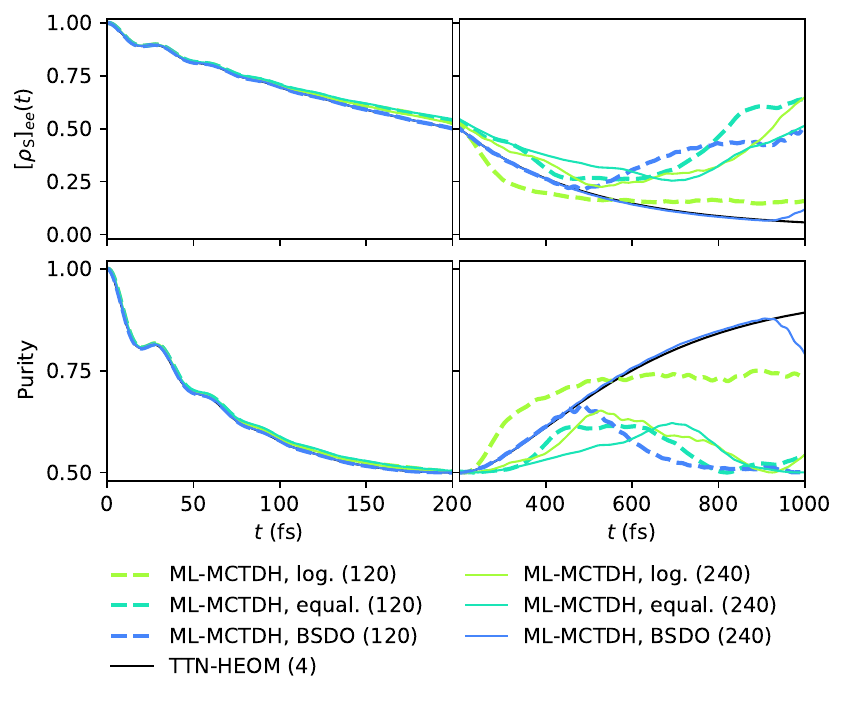}
  \caption{The same dynamics as Fig.~\ref{fig:toy2} but with a larger cutoff frequency $3000~\invcm$ and the number of discretized DoFs is set to be $J=120$ and $J=240$ for the ML-MCTDH computations.}
  \label{fig:toy3}
\end{figure}

If we further increase the cutoff frequency to $3000~\invcm$, the ML-MCTDH with  {the three} discretization strategies do not require the correction from the tail of the spectral density and can capture the correct dynamics up to the  {state of minimal purity}, see Fig.~\ref{fig:toy3}.
However, the ML-MCTDH with both logarithmic and equalized discretization strategies cannot capture the correct dynamics from the   {state of minimal purity} to the thermalized state, even with $J=240$ discretized DoFs.  {Employing the BSDO scheme improves this and yields converged dynamics until $\sim 900$ fs.}
 {For BSDO, the range of validity of the dynamics matches with the range of validity of the BCF decomposition in the time domain as shown in Fig. S3(a, b). In turn, for the equalized and logarithmic discretization, despite they can capture the overall trend of the BCF on the time domain, they are unable to correctly track the whole thermalization because they do not properly sample the tail of the spectral density in the frequency domain. }
This contrast to the TTN-HEOM with $K=4$ which can already capture the correct dynamics  {up to thermalization}.

\subsubsection{Seven-site model for Fenna--Matthews--Olson complex}
We now focus on the dynamics in the photosynthetic Fenna--Matthews--Olson (FMO) complex using the two tensor network methods.
The FMO complex transfers the photoexcitation energy from the antenna complex of green sulfur bacteria to the  photosynthetic reaction center.\cite{Adolphs2006} 
It consists of seven bacteriochlorophyll sites that transfer photoexcitation energy by exciton coupling between them.
The Hamiltonian of the FMO complex can be described by the spin-boson model with the system Hamiltonian \cite{Ishizaki2009a}
\begin{equation} 
  \begin{multlined}
    H_{\text{S}} = \\\begin{pmatrix} 200 & -87.7 & 5.5 & -5.9 & 6.7 & -13.7 & -9.9 \\
      -87.7 & 320 & 30.8 & 8.2 & 0.7 & 11.8 & 4.3\\
      5.5 & 30.8 & 0 & -53.5 & -2.2 & -9.6 & 6\\
      -5.9 & 8.2 & -53.5 & 110 & -70.7 & -17 & -63.3\\
      6.7 & 0.7 & -2.2 & -70.7 & 270 & 71.1 & -1.3\\
      -13.7 & 11.8 & -9.6 & -17 & 71.1 & 420 & 39.7\\
      -9.9 & 4.3 & 6 & -63.3 & -1.3 & 39.7 & 230 \end{pmatrix} \\ (\invcm),
    \end{multlined}
    \end{equation} 
and the system-bath interaction Hamiltonian $H_{\text{SB}}$ is given by $H_{\text{SB}} = \sum_{i} Q^{(i)}_{\text{S}} X^{(i)}_{\text{B}}$,
where $Q^{(i)}_{\text{S}} = \ket{i}\bra{i}$ is the system operator at the $i$-th site for $i = 1,\ \ldots,\ 7$, and $X^{(i)}_{\text{B}} = \sum_{k=1}^{K} g_k^{(i)} \left(a_k^{(i)} + \qtya{a_k^{(i)}}^\dagger\right)$ is the bath coordinate coupled to the $i$-th site.
Here for the FMO complex, the bath at each site is modeled as independent harmonic reservoirs with the same DL spectral density to reduce the number of modeling parameters.
This is a model that is often used in the literature.\cite{Ishizaki2009a,Duan2022,Rodrguez2025,Gestsson2025} 
The $i$-th bath is described by a spectral density 
\begin{equation}
  \mathcal{J}^{(i)}(\omega) =  \sum_{k=1}^{K} \qtya{g_k^{(i)}  \delta(\omega - \omega_k) - g_k^{(i)} \delta(\omega + \omega_k)},
\end{equation}
and the bath for each site is considered to be independent from one another.
The overall Hamiltonian is thus
\begin{equation}
  H = H_{\text{S}} + \sum_{i} H_{\text{B}}^{(i)} + H_{\text{SB}},
\end{equation}
with $H_{\text{B}}^{(i)} = \sum_{k=1}^{K} \omega_k \qtya{a_{k}^{(i)}}^\dagger a_{k}^{(i)}$.

We consider the case that each site $\ket{i}$ is coupled to a DL bath with the same reorganization energy $\lambda_\text{D} = 35~\invcm$ and the same relaxation timescale $\tau=\gamma_\text{D}^{-1}$ at $77$~K.
In this case, in the ML-MCTDH, each Drude bath is discretized by  {60} modes with a cutoff frequency at $1000~\invcm$, which results in a total of  {$420$} modes for all baths.
For the TTN-HEOM, the dissipator from each bath are additive.
For each DL bath, we use Pad\'e $[4/5]$ approximant for the evaluation of the BCF, which gives the number of auxiliary modes $K=6$ for each bath in the HEOM, resulting in a total of $42$ auxiliary modes for all baths.
The dimensions of the Fock basis for each discretized DoF in the ML-MCTDH and for each auxiliary mode in the HEOM are all set to be $5$.
In the simulations, a balanced tree structure with maximal bond threshold $R=32$ was used for both TTN-HEOM and ML-MCTDH.

\begin{figure}[tbp]
    \centering
    \includegraphics[width=1.0\linewidth]{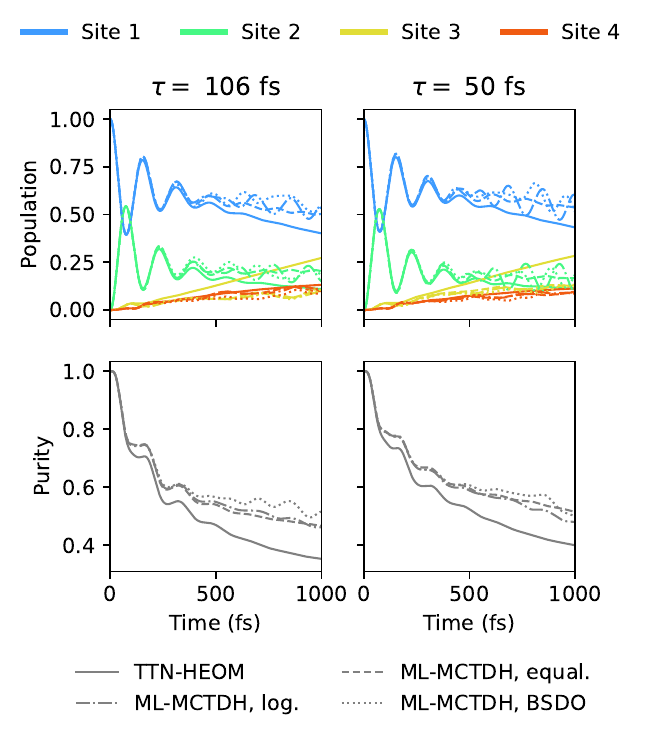}
    \caption{Population and purity dynamics in FMO using ML-MCTDH with different discretization strategies (dashed lines: equivalent reorganization energy discretization strategy; dash-dotted lines: logarithmic discretization {; dotted lines: BSDO discretization}) and TTN-HEOM with implicit discretization based on Pad\'e approximant (solid lines). }
    \label{fig:fmo}
\end{figure}

 {As discussed in Sec.~\ref{sec:ch7-sbm-relaxation}, the validity of the ML-MCTDH with discretization depends on the range of convergence in frequency and time of the BCF decomposition.}
Here we focus on the first one picosecond of the dynamics.
The initial system state is set to be at site $1$ as $\rho_\text{S}(0) = \ket{1}\bra{1}$.
The computed dynamics are shown in Fig.~\ref{fig:fmo}. 
In the right panels we use $\tau=50~\text{fs}$ which corresponds to a faster dissipation process, while in the left panels we use $\tau=106~\text{fs}$ which corresponds to a slower dissipation process.
We compare the results from the explicit discretization with the ML-MCTDH and the implicit discretization with the TTN-HEOM.
The population dynamics of the site $i$ is as $\rho_{ii}(t)$, and the coherence dynamics is measured by the purity $\Tr \rho_\text{S}^2(t)$ of the reduced density matrix $\rho_\text{S}(t)$.
The results show that either tensor network method can capture the overall trend for the first picosecond of the population and decoherence dynamics in the FMO complex.
While the explicit discretization with the ML-MCTDH shows a more oscillatory behavior than the implicit discretization with the TTN-HEOM, especially when the relaxation time $\tau$ is short, the qualitative trend of both population and decoherence dynamics are similar for  {all three discretization} methods.
However, the ML-MCTDH with  {$420$} discretized modes quantitatively deviates from the TTN-HEOM results with $42$ auxiliary modes, especially when the system is approaching the  {state of minimal purity}.

 {As to the computational cost, in Table~\ref{tab:fmo}, we show the space and time complexity of the FMO computations with the Drude--Lorentz bath and $\tau = 50$~fs at 77 K using different methods as implemented in \texttt{TENSO}.
We focus on the first 100 fs of the dynamics where the ML-MCTDH results with all three discretization methods converge with the TTN-HEOM result.
Here since the dimension of the primitive basis for each discretized mode used in the ML-MCTDH (2) is smaller than the one for each quasiparticle in the TTN-HEOM (5), the space complexity (measured by the number of complex floating-point [CFP] numbers needed to store the TTN at 100 fs) is comparable between the TTN-HEOM and the ML-MCTDH.
However, the time complexity for the TTN-HEOM is significantly smaller because it has less core tensors and thus a simpler tree structure than the ML-MCTDH even when the average size  of each core tensor is larger. 
Since the operations in the implemented algorithms in \texttt{TENSO} between different tensors are less optimized compared to the operations of a dense tensor implemented in the industry-standard PyTorch package, the overall time cost in the TTN-HEOM is much smaller (40--70 times) than the ML-MCTDH with these three discretization methods compared to advantages of the memory cost (2--3 times).}

 {It is important to emphasize that the actual CPU time of each method is influenced by many factors such as the choice of TTN tree structure and the details of the propagation algorithms.
Specifically, in our comparison, we employ one balanced binary tree which is automatically generated by  \texttt{TENSO}.
We also use the mixed propagation strategies with initial steps using two-site projector-splitting algorithm with adaptive ranks \cite{Lubich2018,Lindoy2021a,Lindoy2021b}, followed by the direct integration using the decomposed master equations for each tensor in the TTN based on TDVP with the regularization technique \cite{Wang2021}. 
For the integration of each decomposed master equation, we use the RK4(5) method \cite{Shampine1986} which is a Runge--Kutta method of fourth order with an error estimator of fifth order for adaptive time step.}

\begin{table*}
\begin{tabular}{ccccc}
    \toprule
    Method & \multicolumn{3}{c}{ML-MCTDH} &  TTN-HEOM \\
    Discretization & Logarithmic & Equalized & BSDO & (Pad\'e) \\
    \midrule
    \# of modes for each bath & 60 & 60 & 60 & 6 \\
    Time cost (h) & $30$ & $21$ & $37$ & $0.5$ \\ 
    TTN size (CFP) & $104693$ & $158081$ & $92472$ & $43264$\\
    \bottomrule
\end{tabular}
\caption{%
Space and time complexity of the FMO computation with the Drude--Lorentz bath and $\tau = 50$~fs at 77 K using different methods.
The time cost is estimated by the CPU time for simulating the first 100 fs of dynamics using 8 cores of Intel Xeon Gold 6330 CPU \@ 2.00 GHz. 
The TTN size is counted by the number of complex floating-point (CFP) numbers needed to store the TTN at 100 fs.
}
\label{tab:fmo}
\end{table*}

\subsection{ {Brownian Environment}}
\label{sec:ch7-br}

\begin{figure}[h]
    \centering
    \includegraphics[width=\linewidth]{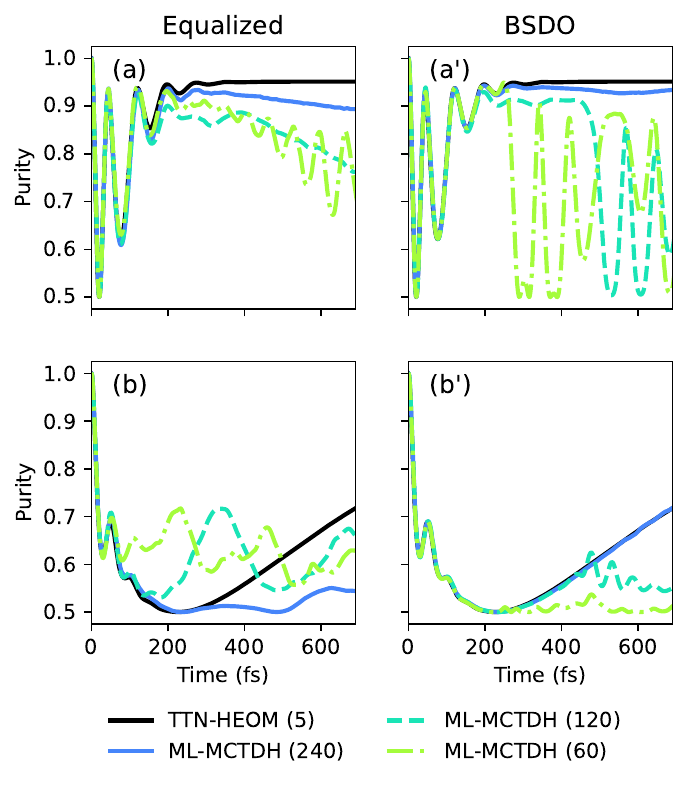}
    \caption{ {Identical computations to those in Fig.~\ref{fig:toy3} but with a Brownian oscillator bath with an effective frequency [(a) and (a')] at resonance $\omega_1 =1000~\invcm$ or [(b) and (b')] out of resonance $\omega_1 =500~\invcm$ with the two-level system. In both cases, the width ${\gamma_\text{B}} = 100~\invcm$ and reorganization energy ${\lambda_\text{B}} =200~\invcm$.
    The ML-MCTDH results are from the equalized [(a) and (b)] or BSDO [(a') and (b')] discretization strategy with cutoff frequency $3000~\invcm$ and with the number of discretized DoFs in the bracket for each label.
    The TTN-HEOM results are from the Pad\'e approximant for the low-temperature corrections with overall $K=5$.}
    }
    \label{fig:b}
\end{figure}

To further test the validity of these observations in a different bath model, we repeated the computation in Fig.~\ref{fig:toy3} but using a Brownian oscillator model of frequency $\omega_1$ with the same lifetime. 
The results are shown in Fig.~\ref{fig:b} using the equalized (a, b) and the BSDO (a', b') discretization strategies. 
The oscillator's frequency $\omega_1$ is chosen at resonance {with $\omega_1 = 1000~\invcm$} (a, a') and off-resonance {with $\omega_1 = 50~\invcm$} (b, b') with the two-level system. 
{In both cases, the width $ \gamma_\text{B} = 100~\invcm$ and reorganization energy $ \lambda_\text{B} =200~\invcm$.}
Figures S1--S3 show the BCF decomposition {with the cutoff frequency $3000~\invcm$} in the time domain.
The observed behavior in the BCF decomposition and the dynamics is qualitatively similar to the one in the presence of Drude-Lorentz bath. That is, for the explicit discretization strategy increasing the number of modes increases the time range in which the strategy can accurately capture the quantum dynamics.  
For the resonant case [Fig.~\ref{fig:b}(a, a')] with 240 discretization modes, this is enough to capture the initial decoherence dynamics to a state that is close to the asymptotic thermal state. However, while the HEOM dynamics remains in the thermal state, the real-time propagation based on the explicit discretization strategy will eventually make the system deviate from this thermal state due to its finite recurrence time. 
In turn, for the non-resonant case [Fig.~\ref{fig:b}(b, b')] the explicit discretization only accurately captures the initial purity decay as the thermalization process is slower than the accuracy range of the approach.
%Xinxian: Is there an issue of convergence in the primitive basis for this type of bath when resonantly coupled. 

\begin{figure}[h]
    \centering
    \includegraphics[width=\linewidth]{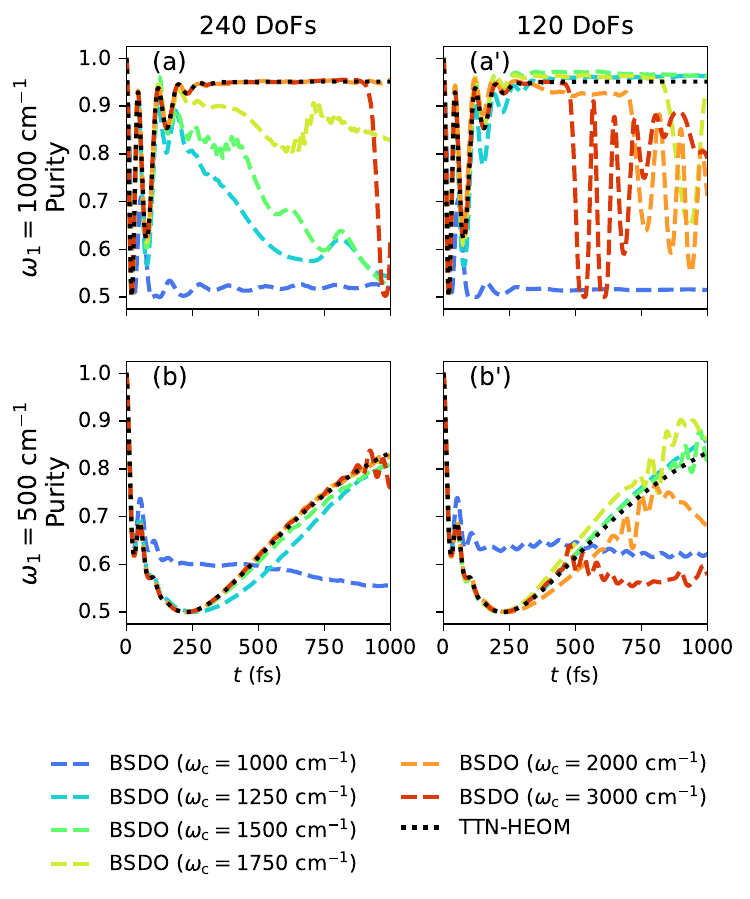}
    \caption{%
    {ML-MCTDH simulation results for the Brownian oscillator bath using BSDO discretizations with different cutoff frequencies.
    (a) and (a') show the results for Brownian oscillator bath with the effective frequency $\omega_1 = 1000~\invcm$, while (b) and (b') with $\omega_1 = 500~\invcm$.
    The number of DoFs used in the BSDO is 240 in (a) and (b), and is 120 in (a') and (b').}
    }
    \label{fig:b-cutoff}
\end{figure}

% Add discussions about BSDO cutoff frequencies
{
It is worth noting that the accuracy of the BSDO method is sensitive to the choice of the frequency-domain interval. 
In Fig.~\ref{fig:b-cutoff} we show the ML-MCTDH simulation results for the Brownian oscillator bath with the resonant frequency $\omega_1 = 1000~\invcm$ [(a) and (a')] and with the off-resonant one $\omega_1 = 500~\invcm$ [(b) and (b')]
using BSDO discretizations with different cutoff frequencies.
We fix the DoFs for the discretization to be 240 [Figs.~\ref{fig:b-cutoff}(a) and (b)] and 120 [Figs.~\ref{fig:b-cutoff}(a') and (b')].
With 240 DoFs, discretizations with a cutoff frequency of $2000~\invcm$ give converged results for the two different Brownian oscillator baths for around $800$ fs of dynamics.
Larger or smaller cutoff frequencies lead to deviations earlier in the dynamics.
%while a smaller cutoff frequency of $1500~\invcm$ or larger cutoff frequency of $3000~\invcm$ can both lead to noticeable deviations for the long-time dynamics.
This behavior is more pronounced for a discretization with 120 DoFs, where a cutoff frequency of $1500~\invcm$ gives better overall trends than those cases with cutoff frequencies $1000~\invcm$ or $2000~\invcm$. 
In turn, the larger cutoff frequency  ($3000~\invcm$) offers more accurate results for early times.
The observed behavior arises because the discretization requires a large enough cutoff frequency to cover all needed modes for long-time dynamics, but cannot be too large such that there are not enough modes around the peak of the bath spectral density.
These results highlight the importance of carefully selecting both the cutoff frequency and the number of discretization degrees of freedom to achieve accurate and converged simulations in the explicit discretization.}

{%
Last, it is known that the Brownian-oscillator spectral density admits an efficient decomposition into a sum of exponentials \cite{Liu2014}, which can make this choice of spectral density more favorable for the HEOM approach from the computational cost perspective. Nevertheless, the HEOM may exhibit numerical instability for this type of bath especially for the weakly damped Brownian oscillators \cite{Dunn2019, Chen2024, Park2024}.}

\section{Conclusion}\label{sec:ch7-con}

In this paper, we have systematically compared explicit and implicit discretization strategies for simulating open quantum dynamics coupled to a dissipative bath focusing on  wavefunction-based time-dependent Schr\"odinger equation (TDSE) approaches and density matrix-based hierarchical equations of motion (HEOM) frameworks. Tensor network techniques, such as TTN-HEOM and ML-MCTDH, further enhanced the scalability of both approaches with the number of features or discretized modes in the bath. 
 {%
We investigated different types of quantum environments that have finite correlation time and can lead to thermalization of the system, including the Drude--Lorentz environment, and the Brownian environments both at resonance and off-resonance with the system. 
We considered three widely used explicit discretization strategies of the bath spectral density---logarithmic, equalized and the bath spectral density orthogonal method.}

Our analysis reveals that the implicit discretization strategy inherent to the HEOM formalism, which relies on structured decompositions of the bath correlation function (BCF), offers significant computational advantages for capturing dissipative dynamics compared to explicit discretization of the bath into discrete harmonic modes. This efficiency comes from the HEOM's ability to encode the bath's memory effects through a hierarchy of auxiliary density matrices, capturing the dissipative and irreversible nature of the bath while retaining numerical exactness.
In particular, HEOM-based methods excel for baths with rapidly decaying BCFs, where a small number of auxiliary modes suffices to reproduce dissipative behavior.   {The unitary wavefunction-based method with finite discretized modes can only mimic the thermalization up to a certain time.} This explicit discretization requires a large number of modes to approximate continuum baths, leading to prohibitive dimensionality even with advanced tensor compression  {and advanced discretization techniques}.

We also note that although the HEOM-based methods are more efficient for dissipative dynamics,  {it is known that the numerical propagation of HEOM can become numerically unstable \cite{Dunn2019,Li2022,Krug2023,Park2024} when the bath correlation functions are long-lived.
By contrast, wavefunction-based ML-MCTDH with unitary propagator \cite{Haegeman2016,Kloss2017,Dunnett2021} can preserve the stability of the time-dependent Schr\"odinger equations in general.}
 {Recently, it was reported that the explicit and implicit strategies can be also unified in a pseudomode theory with a quasi-Lindblad master equation framework that can achieve improved numerical stability with respect to the HEOM \cite{Xu2023, Park2024, Lindoy2025}. Future prospects include contrasting implicit, explicit and mixed discretization strategies for environments with long-lived correlations.}

\section*{Supplementary Material}
See the Supplementary Material for plots of the discretized bath correlation function in the time domain, and a convergence study of the ML-MCTDH computations with the number of primitive basis.

\section*{Data Availability}
The data that support the findings of this study are available from the corresponding author upon reasonable request.

\begin{acknowledgments}
  This material is based on work supported by the U.S.\ Department of Energy, Office of Science, Office of Basic Energy Sciences, Quantum Information Science Research in Chemical Sciences, Geosciences, and Biosciences Program under Award No.~DE-SC0025334.
\end{acknowledgments}

\bibliography{references.bib}

\clearpage % 
\pagenumbering{gobble}
\thispagestyle{empty}
\begin{figure*}
    \centering
    \includegraphics[page=1,width=\linewidth]{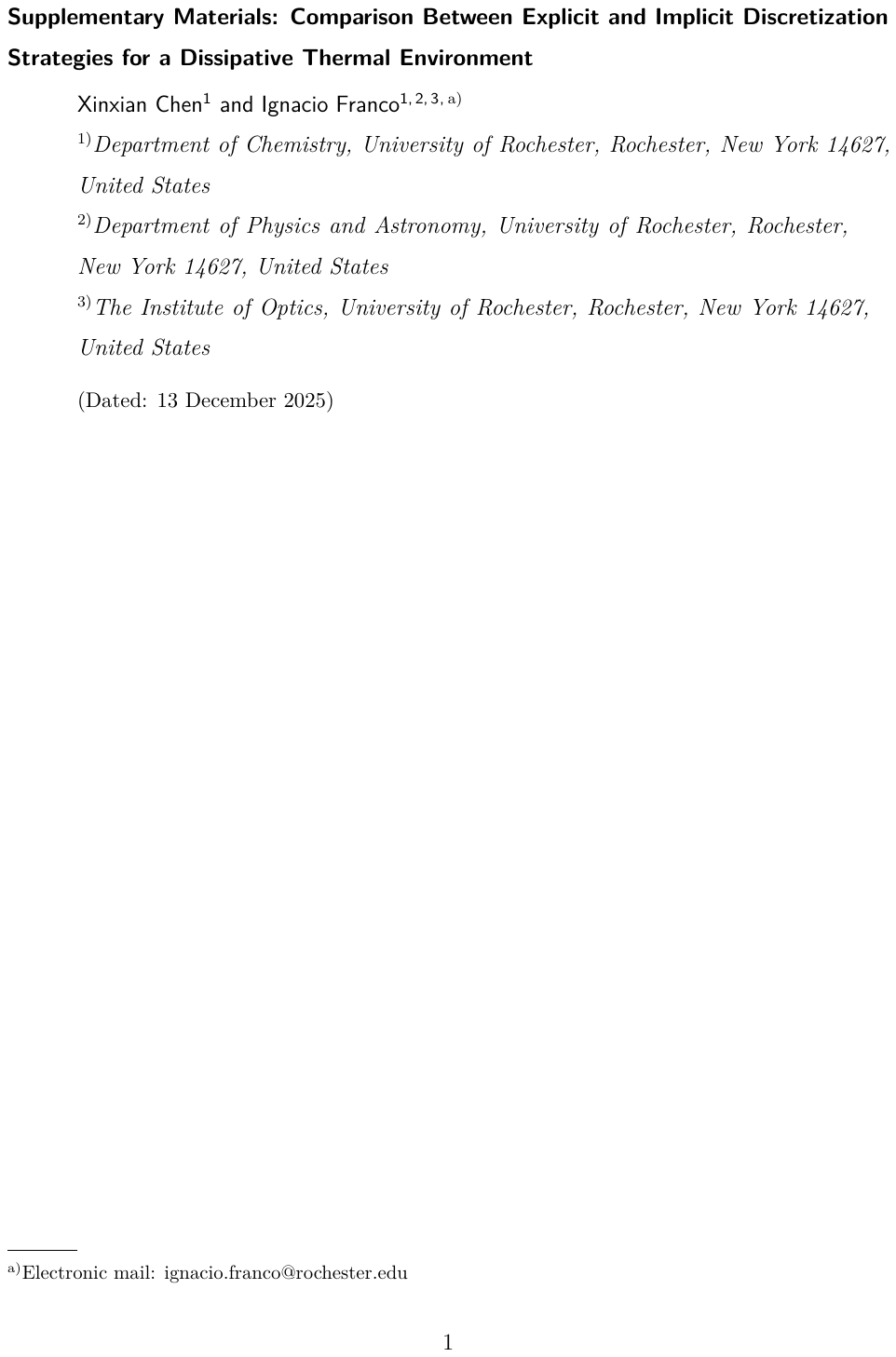} %
\end{figure*}
\thispagestyle{empty}
\begin{figure*}
    \centering
    \includegraphics[page=2,width=\linewidth]{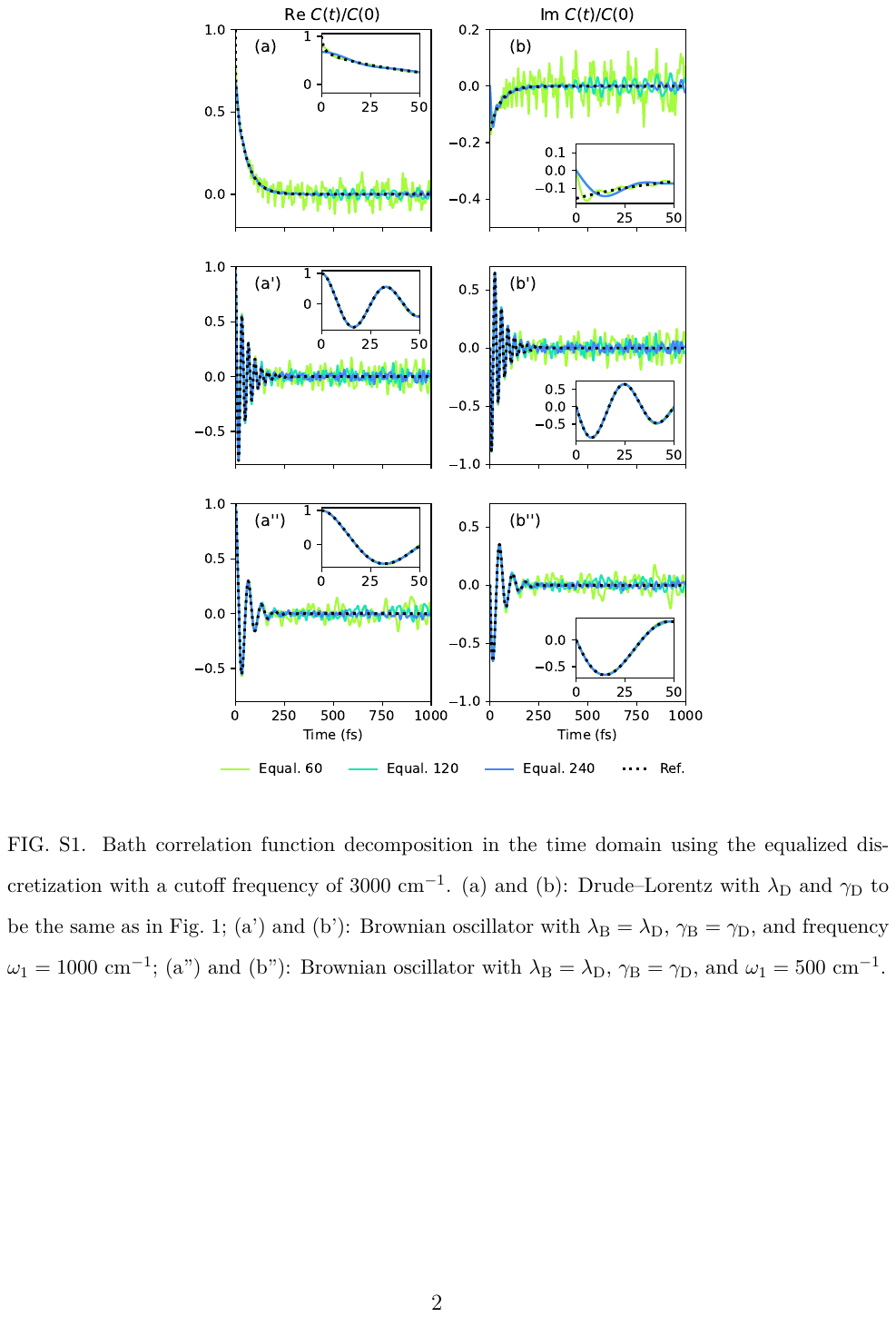} %
\end{figure*}
\thispagestyle{empty}
\begin{figure*}
    \centering
    \includegraphics[page=3,width=\linewidth]{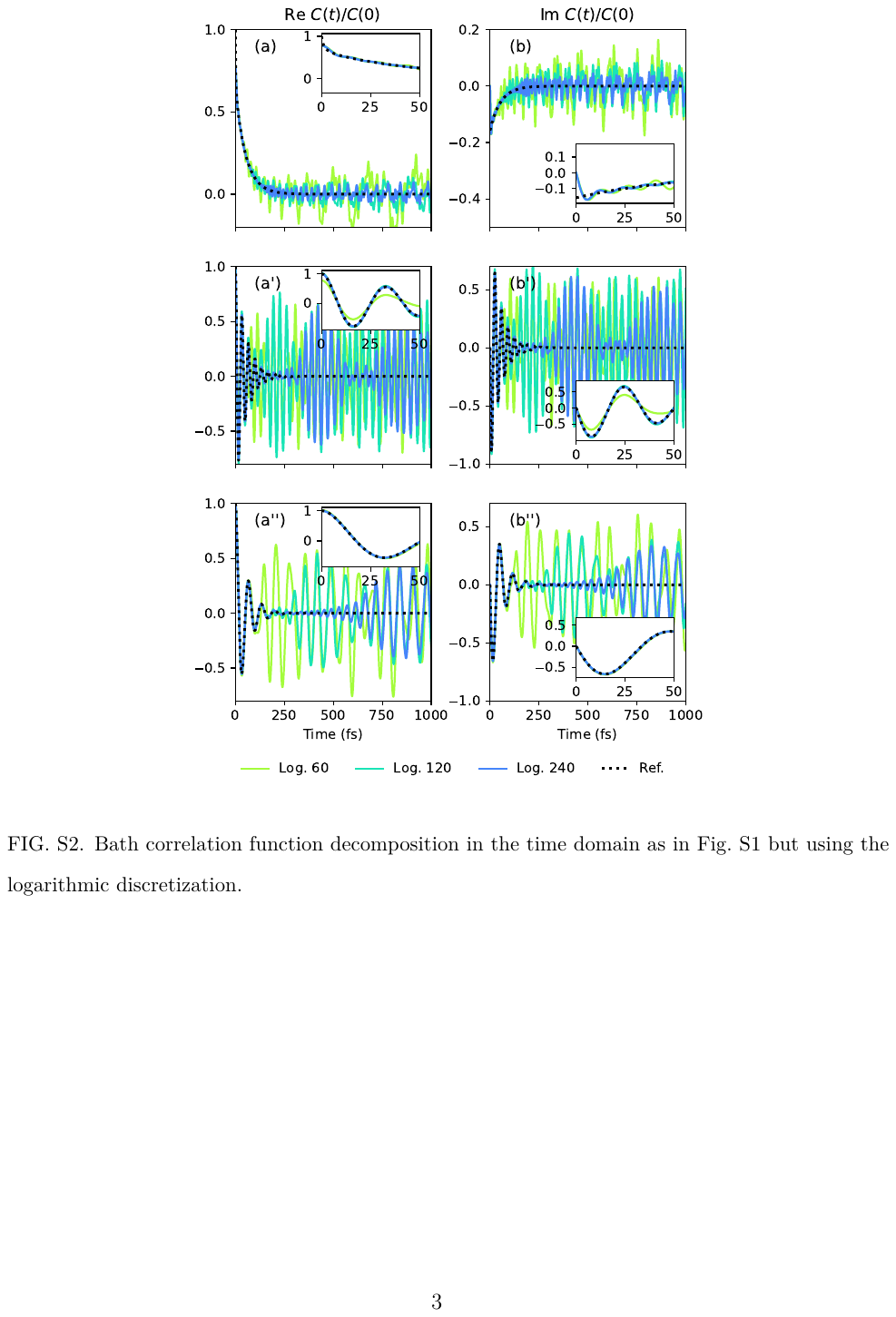} %
\end{figure*}
\thispagestyle{empty}
\begin{figure*}
    \centering
    \includegraphics[page=4,width=\linewidth]{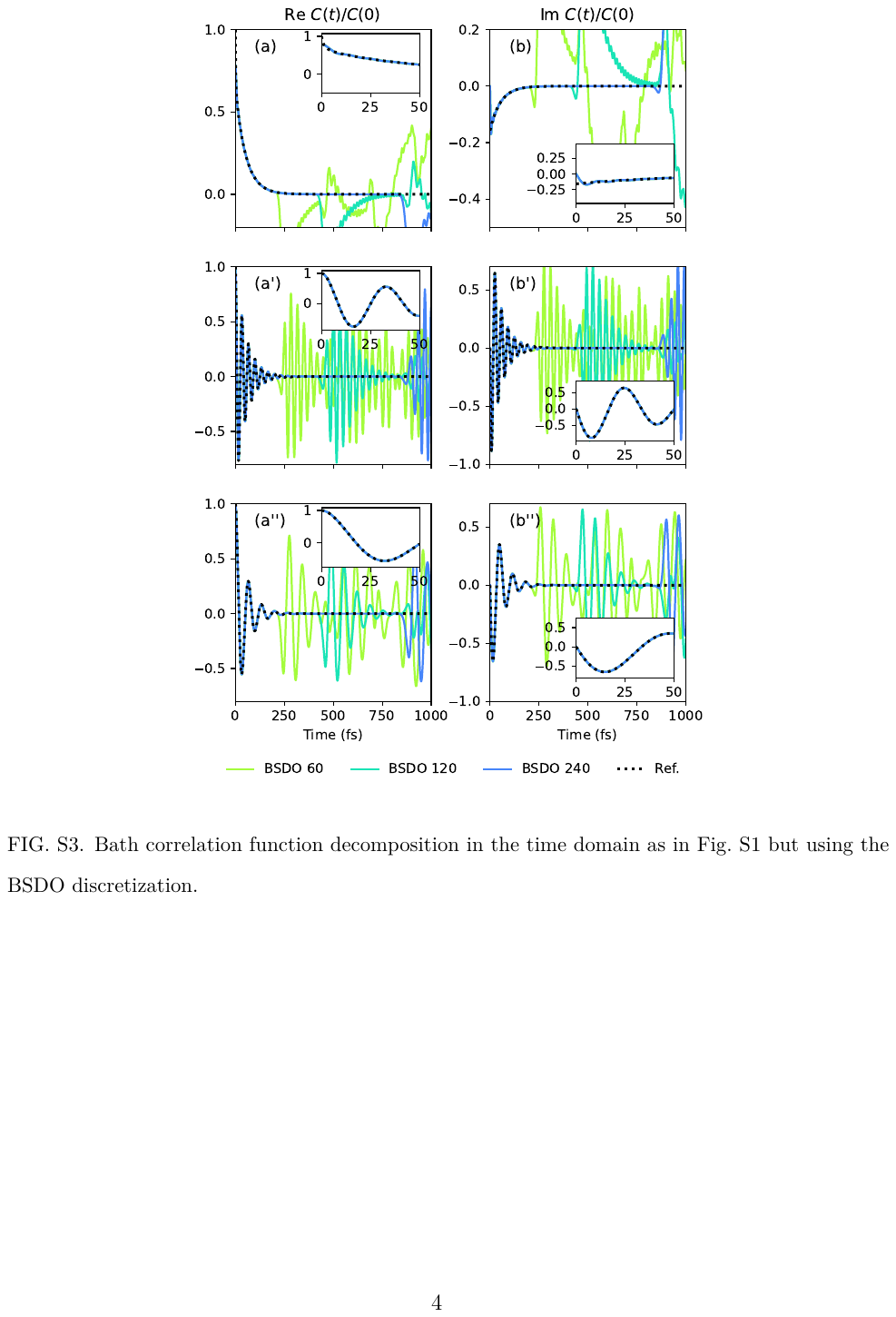} %
\end{figure*}
\thispagestyle{empty}
\begin{figure*}
    \centering
    \includegraphics[page=5,width=\linewidth]{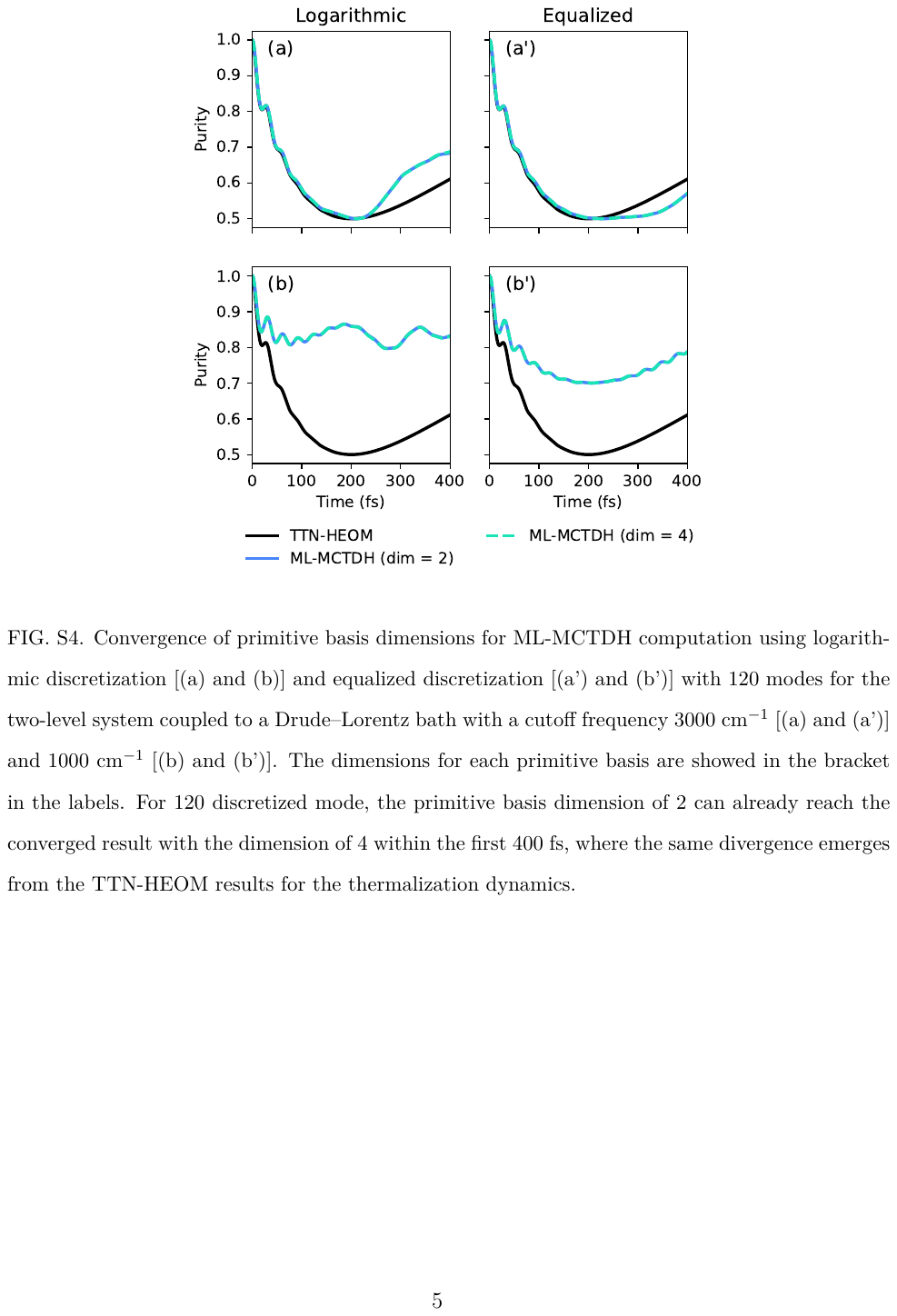} %
\end{figure*}

\end{document}